\documentclass{JHEP}
\usepackage{graphicx}
\usepackage{epsfig}
\usepackage{amssymb}
\usepackage{amsmath}
\usepackage{float}

\newcommand{\bea}{\begin{eqnarray}}
\newcommand{\eea}{\end{eqnarray}}
\newcommand{\bean}{\begin{eqnarray*}}
\newcommand{\eean}{\end{eqnarray*}}
\newcommand{\nn}{\nonumber \\}

\def\W #1{\widetilde{#1}}

\def\Tr{\mathop{\rm Tr}}
\def\det{\mathop{\rm det}}

\def\eref#1{(\ref{#1})}

\def\a{{\alpha}}

\def\b{{\beta}}

\def\eps{\epsilon}

\def\Sl{\sum\limits}
\def\Label#1{\label{#1}%
  \smash{\hbox to0pt{\raise1ex\hbox{\tiny[#1]}\hss}}}

\allowdisplaybreaks
\title{The Construction of Dual-trace Factor in Yang-Mills Theory}

\author{Yi-Jian Du ${}^a$, Bo Feng ${}^{b,c}$, Chih-Hao Fu ${}^d$
~~~~~~~~~~~~~~\\  ${}^a$ Department of Physics and Center for Field Theory
and Particle
 Physics, Fudan University, Shanghai 200433, P.R China\\${}^b$ Zhejiang Institute of Modern Physics, Zhejiang
University, 38 Zheda Road Hangzhou, 310027 P.R China\\${}^c$
Center of Mathematical Science, Zhejiang
University, 38 Zheda Road Hangzhou, 310027 P.R China\\
${}^d$Department of Electrophysics, National Chiao-Tung University and Physics Division,
National Center for Theoretical Sciences, Hsinchu, Taiwan, R.O.C.
~~~~~~~\\
 \email{ yjdu@fudan.edu.cn; b.feng@cms.zju.edu.cn; zhihaofu@nctu.edu.tw \hskip0.5cm} }

\date{\today}
\abstract{Recently,
a BCJ dual of the color-ordered formula for Yang-Mills amplitude was
proposed, where the dual-trace factor satisfies cyclic symmetry and
KK-relation. In this paper, we present a systematic construction of
the dual-trace factor based on its proposed relations to kinematic
numerators  in dual-DDM form. We show that the construction
presented
respects relabeling symmetry. In addition, we show that using
relabeling symmetry as  conditions, the same construction can be
solved independently.

}


\begin{document}
\section{Introduction}

In recent years, a significant progress in the study of scattering amplitudes is the discovery of
 color-kinematic duality\cite{Bern:2008qj}.
At tree-level, the duality states that the complete Yang-Mills tree amplitude ${\cal A}_{tot}$ can always be written
into the following  formula
\bea {\cal A}_{tot} = \sum_i { c_i n_i\over
D_i},  ~~~\label{BCJ-form}
\eea
where the sum runs over all distinct cubic tree diagrams. In the
formulation of Bern, Carrasco and Johansson (BCJ),
the  kinematic factors $n_i$, which we will call ``BCJ numerator'', satisfy the same algebraic relations
as those  of the color factors $c_i$, i.e.,
\bea
{\rm antisymmetry}:&~~& c_i\rightarrow-c_i\Rightarrow n_i\rightarrow-n_i\nn
{\rm Jacobi-like~identity}:&~~& c_i+c_j+c_k=0\Rightarrow n_i+n_j+n_k=0.~~\label{BCJ-duality}
\eea

The duality between   color and kinematic factors provides
 strong constraints on color-ordered
 Yang-Mills tree amplitudes.
Specifically,
the antisymmetry of kinematic
factors implies Kleiss-Kuijf \cite{Kleiss:1988ne, DelDuca:1999rs}
relations, while the Jacobi-like identity implies BCJ
relations\cite{Bern:2008qj}. The newly discovered BCJ relations have been
understood both from  string
\cite{BjerrumBohr:2009rd,Stieberger:2009hq,Tye:2010dd} and field
theory \cite{Feng:2010my,Jia:2010nz,Tye:2010kg,Chen:2011jxa} perspectives. BCJ
relations
  also serve as
  the key to the understanding of KLT relations \cite{KLT}, which express  gravity tree amplitude
in terms of products of two color-ordered Yang-Mills tree amplitudes
(See
\cite{BjerrumBohr:2010ta,BjerrumBohr:2010zb,BjerrumBohr:2010yc,Feng:2010br}).
Although a proof at loop-levels is currently absent,
explicit calculations show that the duality \eqref{BCJ-duality} is
also satisfied at the first few loops
\cite{Bern:2010ue,Bern:2012uf,Bern:2011rj,BoucherVeronneau:2011qv,Naculich:2011my,
Oxburgh:2012zr,Carrasco:2012ca,Boels:2013bi,Bjerrum-Bohr:2013iza,Bern:2013yya}.

Because of the  applications  in the study of  Yang-Mills
 and gravity amplitudes, the  construction of  BCJ numerators has become
 an important problem, and there are many discussions in the literature.
In \cite{Mafra:2011kj}, BCJ numerators were constructed by string
pure-spinor method. In \cite{Monteiro:2011pc,BjerrumBohr:2012mg}, a
light-cone gauge approach for the kinematic algebra was suggested,
which provides a natural algebraic explanation to BCJ duality. Using
this approach  BCJ numerators of MHV Yang-Mills amplitude  and all
amplitudes in self-dual Yang-Mills theory can indeed be expressed as
structure constants of a  diffeomorphism algebra
\cite{Monteiro:2011pc,BjerrumBohr:2012mg}. Based on this idea, we
have  proposed a more general kinematic algebra in \cite{Fu:2012uy},
from which one can construct BCJ numerators at tree-level in
arbitrary D-dimensions and for arbitrary helicity configurations.


The fact that
the color factors $c_i$ and the kinematic numerators $n_i$ share the same algebraic structure,
also suggests that the existing decompositions of Yang-Mills tree
amplitudes may have  color-kinematic counterparts.
Traditionally, we have two different decompositions of Yang-Mills
amplitudes\cite{DelDuca:1999rs}
\bea
{\rm Trace~form}: &~~~ &{\cal A}_{tot}  =g^{n-2} \sum_{\sigma\in
S_{n-1}}{\rm Tr} (T^{1}... T^{\sigma_n})
A(1,\sigma_2,...,\sigma_n),~~~\label{Trace-form}\\
{\rm DDM~form}: &~~~ & {\cal A}_{tot}  = g^{n-2} \sum_{ \sigma\in S_{n-2}}
c_{1|\sigma(2,..,n-1)|n} A(1,\sigma,n),~~~\label{DDM-form}
\eea
where $g$ is the coupling constant,   and $A$'s  are the
color-ordered amplitudes. In Trace form the generator $T^a$ is given
by fundamental representation of $U(N)$ group, while in DDM form
$c_{1|\sigma(2,..,n-1)|n} $ is constructed using structure constants
$f^{abc}$ as
\bea c_{1|\sigma(2,..,n-1)|n}=f^{1 \sigma_2 x_1} f^{x_1
\sigma_3 x_2}... f^{x_{n-3} \sigma_{n-1} n}~.~~\label{DDM-c}\eea
The equivalence between BCJ form \eqref{BCJ-form} and DDM
form
 was shown in \cite{Du:2011js},
 where both forms were proven to be equivalent to
 the KLT relation of color ordered scalar theory. To show
 the equivalence
between DDM form and  Trace form  \cite{DelDuca:1999rs},
 the following two properties of $U(N)$ Lie algebra
are essential
\bea {\rm Property~One:} & ~~~ & (f^a)_{ij}= f^{aij}={\rm Tr}( T^a[
T^i,
T^j]),~~~\label{group-1}\\
{\rm Property~Two:} & ~~~ &  \sum_a {\rm Tr}( X T^a) {\rm Tr}(T^a
Y)= {\rm Tr}( XY)\nn
& & \sum_{a} {\rm Tr}( X T^a Y T^b)={\rm Tr}( X){\rm Tr}( Y)~.~\label{group-2}\eea
Based on  BCJ duality \eqref{BCJ-form}, it is natural to exchange
the roles of $c_i$ and $n_i$ and consider the following two dual
forms
\bea {\rm Dual~Trace~form}: &~~~ &{\cal A}_{tot}  = g^{n-2}\sum_{\sigma\in
S_{n-1}} \tau_{1\sigma_2... \sigma_n}
\W A(1,\sigma_2,...,\sigma_n),~~~\label{Dual-Trace-form}\\
{\rm Dual~DDM~form}: &~~~ & {\cal A}_{tot}  = g^{n-2} \sum_{ \sigma\in
S_{n-2}} n_{1|\sigma(2,..,n-1)|n}\W
A(1,\sigma,n),~~~\label{dual-DDM-form} \eea
where $\W A$'s are color ordered tree amplitudes of scalar theory with
$f^{abc}$ as its cubic coupling constants (see  references
\cite{Bern:1999bx, Du:2011js}) and $\tau$ (which we will call
``Dual-trace factor'' or simply $\tau$-function) is required to be
cyclic invariant. Indeed, the Dual-DDM form was given in
\cite{Bern:2010yg} while the Dual-Trace form was conjectured in
\cite{Bern:2011ia} with explicit constructions for the first few
lower-point amplitudes and a general construction was suggested in
\cite{BjerrumBohr:2012mg}.

 Although the existence of the above two dual forms were established,
a systematic Feynman rule-like prescription to
$\tau$-functions and BCJ numerators $n_\sigma$ is not yet
 known at this moment. The dual DDM-form 
was studied in \cite{Fu:2012uy}, where the construction of
BCJ numerators  $n_{1|\sigma(2,..,n-1)|n}$ was 
given (see (\ref{An-DDM})). Although the result in  \cite{Fu:2012uy}
for BCJ numerators is just a small step towards the systematic local
diagram construction, it does give us some useful applications.


In this paper, we  use BCJ numerators to  systematically
construct  the dual-trace factor $\tau$ and realize the proposed
Dual-trace form \eref{Dual-Trace-form}. Unlike the trace factor ${\rm Tr}(T^a...)$
which satisfies  cyclic symmetry by construction,
there is no  relation presumed among these $\tau$-functions, thus in principle there
are $n!$ $\tau$-functions we need to determine. Any solution to
these $n!$ $\tau$-functions will be a rightful choice as long as it gives
the right total amplitude \eref{Dual-Trace-form}. However, from the
way dual-DDM forms are labeled, we see that there are only $(n-2)!$ BCJ numerators
$n_{1\sigma n}$
 needed to completely fix the total amplitude, thus it is
very natural to impose some relations  to reduce the
 number of independent $\tau$'s. These  imposed relations are
\cite{Bern:2011ia}:
\begin{itemize}
\item (A) {\bf Cyclic symmetry:}
\bea \tau_{12...n}=\tau_{n1...(n-1)}.~~\label{cyclic-symmetry} \eea
Using the cyclic symmetry, we can fix the first index to be any particular number, for example,
$1$, thus the number of independent $\tau$-functions is reduced to $(n-1)!$.

\item (B) {\bf KK-relation:}
\bea \tau_{1,\alpha,n,{\beta}^T}=(-1)^{n_{\beta}}\Sl_{\{\sigma\}\in
OP(\{\alpha\}\bigcup\{\beta\})}\tau_{1,\sigma,n},~~\label{KK-relation}
\eea
where $n_\b$ is the number of elements in the set $\b$, and $\b^T$
denotes the inverse of the ordering of set $\b$. The sum in
\eref{KK-relation} is over all permutations of the set
$\{\alpha\}\bigcup\{\beta\}$ where relative ordering inside both
subsets $\a$ and $\b$ are kept.
Using this relation, we can fix two particular numbers, for example, $1$ and $n$,
at the first and last positions, thus the number of
independent $\tau$-functions is $(n-2)!$.

\end{itemize} Having imposed the above two relations, we need to find these $(n-2)!$
independent $\tau$-functions, such that  relation
\eref{Dual-Trace-form} is satisfied. This problem is solved by
imposing the following
 relations between $(n-2)!$ BCJ numerators  $n_{1\sigma n}$\footnote{Since in our whole paper, we will only
use BCJ numerators  of the DDM-chain form, to simplify the notation we will use
$n_{123...n}\equiv n_{1|23...|n}$. } and $(n-2)!$ $\tau_{1\sigma n}$-functions
 (where $\sigma\in S_{n-2}(23...(n-1))$)
\bea n_{1\sigma_2...\sigma_{(n-1)}n}=\tau_{1[\sigma_2,[...,[\sigma_{n-1},n]...]]},
~~\label{n-tau-relation} \eea
here $[~,~]$ denotes the antisymmetric combination, for example
$n_{123}=\tau_{1[2,3]}=\tau_{123}-\tau_{132}$.

After solving $\tau_{1\sigma n}$ as linear combinations of
$n_{1\sigma n}$ using \eref{n-tau-relation}, we can use
\eref{KK-relation} and \eref{cyclic-symmetry} to obtain all other
$\tau$-functions. The claim
is that  if we put these $\tau$'s back
to the right hand side of \eref{Dual-Trace-form},  we do get
the left hand side.
In fact the proof of equivalence of
Trace form \eref{Trace-form} and   DDM form \eref{DDM-form} 
only 
relies on two facts: (1) Partial amplitudes $A(1,\sigma,n)$ are
cyclic symmetric and satisfy KK-relations. (2) We have two different  factors
(in this case the
 trace  and the color factor)
satisfying the relation
\bea
c_{1\sigma_2...\sigma_{(n-1)}n}=\Tr(T^{1}[T^{\sigma_2},[...,[T^{\sigma_{n-1}},T^{n}]...]]).
\label{c-tau-relation}\eea
Noting that the imposed relation \eref{n-tau-relation} is exactly
the same as \eref{c-tau-relation}, and the fact that partial amplitudes $\W
A(1,\sigma,n)$ in \eref{Dual-Trace-form} and \eref{dual-DDM-form}
are cyclic symmetric and satisfy KK-relations, we see that the
claim is true.

The above logic is perfectly right, but there are two unclear
points. The first  is that each solution is based on
\eref{n-tau-relation} where a pair of numbers
($(1,n)$ in the example above)
are fixed. Choosing
a different pair  will result in a different solution in principle. For
example, the same $\tau_{123..n}$-function can have two different
expressions corresponding to two different choices of the fixed pair $(i_1,j_1)$ and
$(i_2,j_2)$. Secondly, for a given  
fixed pair, $(1,n)$ for example,
 are the expressions for
 any two $\tau$-functions  
 related to each
other by a corresponding relabeling? 
Specifically,
 let us assume two $\tau$-functions
are $\tau_{\sigma_1}=\sum_{i=1}^{(n-2)!} c_i n_{1\a_i n}$ and
$\tau_{\sigma_2}=\sum_{j=1}^{(n-2)!} d_j n_{1\a_j n}$,
where  $n_{1\a n}$'s are the $(n-2)!$ independent numerators in  KK-basis with $(1,n)$ fixed at  two ends.
If two orderings are related to each other by
a  permutation $P$ of $n$-elements, i.e., $\sigma_2=P(\sigma_1)$,
we can get another expression $ \sum_{i=1}^{(n-2)!} c_i
n_{P(1\a_i n)}$ by relabeling from the expression of $\tau_{\sigma_1}$. Then the question
is that whether we have
\bea \sum_{j=1}^{(n-2)!} d_j n_{1\a_j n}\stackrel{?}{=} \sum_{i=1}^{(n-2)!} c_i
n_{P(1\a_i n)}~.~~~\Label{relabel-property}\eea

These two points are related  to each other. In fact, if
\eref{relabel-property} is satisfied, it should be expected that
$\tau$-functions are unique no matter which pair is taken fixed in
\eref{n-tau-relation} to get them, since different choices can be
related to each other by a permutation. In this paper, we will show
that the solution obtained by our algorithm based on
\eref{n-tau-relation} will have this {\bf natural relabeling property}. In
other words, 
conditions \eref{n-tau-relation} and
natural relabeling property are in fact
consistent with each other. With this understanding we  
present another algorithm that uses
relabeling property to solve $\tau$-functions.

The structure of this paper is the following.  In section
\ref{sec:rev}, we provide a short  review of the kinematic algebra
proposed in \cite{Fu:2012uy}. Then we provide an algorithm to
construct dual-trace factors in section \ref{sec:const-ntr}.
To demonstrate the idea outlined in section \ref{sec:const-ntr}, we  present several
examples in section \ref{sec:example}.
We  discuss the natural
relabeling property and prove that the solution obtained from our algorithm in section \ref{sec:const-ntr}
 does satisfy the relabeling symmetry.
A short summary of this work is given in section \ref{sec:concln}.
Finally, details of the proof of the relabeling property are given
in the Appendix.

\section{Useful properties of BCJ numerators}
\label{sec:rev}
Before presenting  the construction of dual-trace factors, let us review
some useful properties of BCJ numerators discussed  in
\cite{Fu:2012uy}. Especially we will use the Jacobi identity to
establish some relations among BCJ numerators $n_{\a}$ with different orderings. To show
these relations we will follow
the method given in \cite{DelDuca:1999rs}, but there is a small difference.
The proof done in  \cite{DelDuca:1999rs} is for color factors $c_{\sigma}$
constructed using  structure constant $f^{abc}$ of $U(N)$ Lie algebra. The
construction is local in the sense that it is given by a chain-shaped Feynman diagram
with a set of Feynman rules prescribing the contribution of each vertex.
For BCJ numerator $n_\sigma$,  there is still no local construction based on Feynman diagram
with Feynman-like rules prescribing its vertices. In fact, finding a such local construction
is one motivation of our work \cite{Fu:2012uy}, where some progress has been made
towards this goal, which we  review in this section.

To construct local expressions of BCJ numerators $n_\a$,
we need to use  structure constants of kinematic Lie algebra with
the generator given by
\bea
T^{k,a}\equiv e^{ik\cdot x}\partial_{a}.\label{eq:generator}
\eea
Using these generators, we can calculate  commutation relations and find the
  kinematic structure constant
$f^{ab}{}_{c}$
\begin{eqnarray}
[T^{k_{1},a},T^{k_{2},b}] & = & (-i)(\delta_{a}{}^{c}k_{1b}-\delta_{b}{}^{c}k_{2a})\,
e^{i(k_{1}+k_{2})\cdot x}\partial_{c} \nn
 & = & f^{(k_{1},a),(k_{2},b)}{}_{(k_{1}+k_{2},c)}\, T^{(k_{1}+k_{2},c)}. \label{f-def}
\end{eqnarray}
These kinematic structure constants satisfy antisymmetry property
\bea f^{12}_{~~~3}=-f^{21}_{~~~3},~~~\label{Anti} \eea
and Jacobi identity
\begin{equation}
f^{1_{a},2_{b}}{}_{(1+2)^{e}}f^{(1+2)_{e},3_{c}}{}_{(1+2+3)^{d}}
+f^{2_{b},3_{c}}{}_{(2+3)^{e}}f^{(2+3)_{e},1_{a}}{}_{(1+2+3)^{d}}
+f^{3_{c},1_{a}}{}_{(1+3)^{e}}f^{(1+3)_{e},2_{b}}{}_{(1+2+3)^{d}}=0.~~~\label{Jacobi}
\end{equation}
It is worth  noticing that unlike  structure constants of group Lie algebra,
for the $f^{12}{}_{3}$ given in \eref{f-def}  there is no natural way to lift index
$3$ up thus indices $1,2,3$ cannot  be put on the same footing. To distinguish these
three indices, we  use arrows and   the Jacobi identity  \eref{Jacobi} can be
represented as the cyclic sum over three incoming
 arrow legs illustrated in Figure \ref{fig-Jacobi}.

\FIGURE[ht]{\epsfig{file=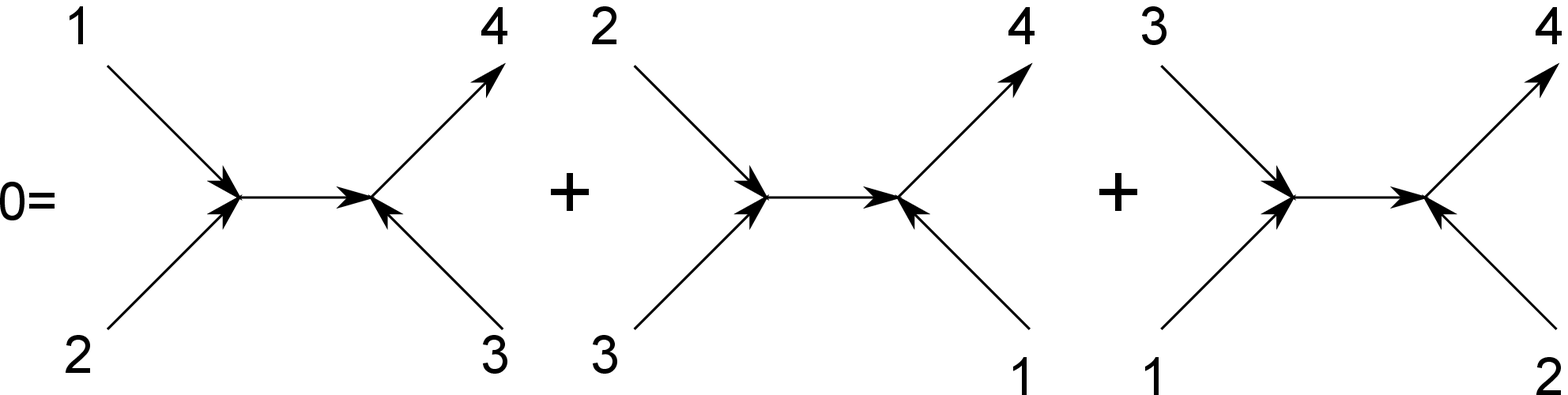,width=13cm}\caption{The Jacobi
identity \eref{Jacobi} of kinematic structure constants can be represented by
the  sum over cyclic orderings of three incoming arrows $1,2,3$.}~~\label{fig-Jacobi}}

Using the above kinematic algebra, we have shown in 
previous work \cite{Fu:2012uy}
 that the total tree-level amplitude of YM-theory can be written as dual-DDM form \eref{dual-DDM-form}
where the {\bf BCJ numerator } is given by
\bea n_{12...(n-1)n}=\sum_{j=1}^N c_j\eps(q_j)\cdot \left(
\begin{array}{c}
 \includegraphics[width=3cm]{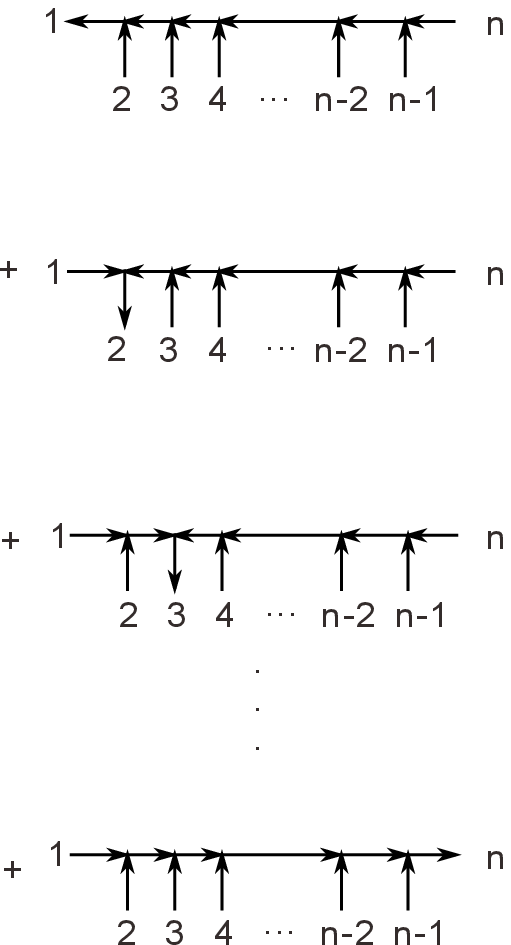} \\ \vspace{-1.5cm} \end{array} \right)~~~\label{An-DDM}\eea
~\\~\\ where each term in the bracket is constructed using kinematic
structure constants as coupling for each cubic vertex. The
$\eps(q_j)$ is defined as $\prod_{t=1}^n \eps_{t}^{\mu_t}(q_{tj})$
 where $ \eps_{t}^{\mu_t}(q_{tj})$ is the polarization vector of the $t$-th external
 particle with gauge choice $q_{tj}$. The $c_j$'s are  coefficients solved by
 our averaging procedure given in \cite{Fu:2012uy}. The explicit expressions
 of $c_j$'s are not important for our purpose here and the only useful fact we need
 is that $c_j$'s are {\sl independent of   color orderings}. In other words,
 for all color orderings of $n_{\a}$, the $c_j$'s are same.
 Because of this structure, when we discuss BCJ numerators, the $\sum_j c_j \eps(q_j)$
 part can be neglected and we will focus only on the part inside the bracket.

Now the part inside the bracket has a Feynman diagram-like structure, much like
 the color structure discussed in \cite{DelDuca:1999rs}. Using
  the same method given in \cite{DelDuca:1999rs} ( i.e., using  the Jacobi identity \eref{Jacobi} and antisymmetry  property
\eref{Anti}), we can find some nice
 relations among BCJ numerators $n_\a$ with different orderings.
For example, we   have the following two identities
\bea &&n_{n\alpha_1\dots\alpha_i1\rho_{1}\dots\rho_{j}(n-1)}=\Sl_{\{\rho\}\in
OP(\{\beta\}\bigcup \{\gamma\})}
(-1)^{r+1}n_{n \a_1...\a_i\gamma_1...\gamma_s (n-1)\b_r...\b_11}.
~~\label{DDM-relations-1} \eea
and
\bea &&n_{n\alpha_1\dots\alpha_i1\rho_{1}\dots\rho_{j}(n-1)}=\Sl_{\{\rho\} \in
OP(\{\beta\}\bigcup \{\gamma\})}
(-1)^{i+s}n_{1\beta_1\dots\beta_r(n-1)\gamma_s\dots\gamma_1\alpha_i\dots\alpha_1n}.
~~\label{DDM-relations} \eea
From now, without explicit explanations, all sets are ordered in all
manipulations, thus
$OP(\{\beta\}\bigcup \{\gamma\})$ denotes all possible unions of two
sets with arbitrary relative ordering between them, but relative
ordering inside each set has been kept (see the explanation after
equation (\ref{KK-relation})). The sum in \eref{DDM-relations-1}
 and  \eref{DDM-relations}  can alternatively be regarded as
  over all possible splittings
of the ordered set $\{\rho\}$ into two ordered subsets $\{\b\}$ and $\{\gamma\}$ (both sets $\{\b\}$ and
$\{\gamma\}$ can be empty set). In each ordered subset, the relative ordering
must be the same as the relative ordering in the mother set $\{\rho\}$.
These two identities are relabeling invariant, i.e., if we act a
 permutation $P\in S_n$ on  $n$-indices on both sides, there two identities still hold.

\FIGURE[ht]{\epsfig{file=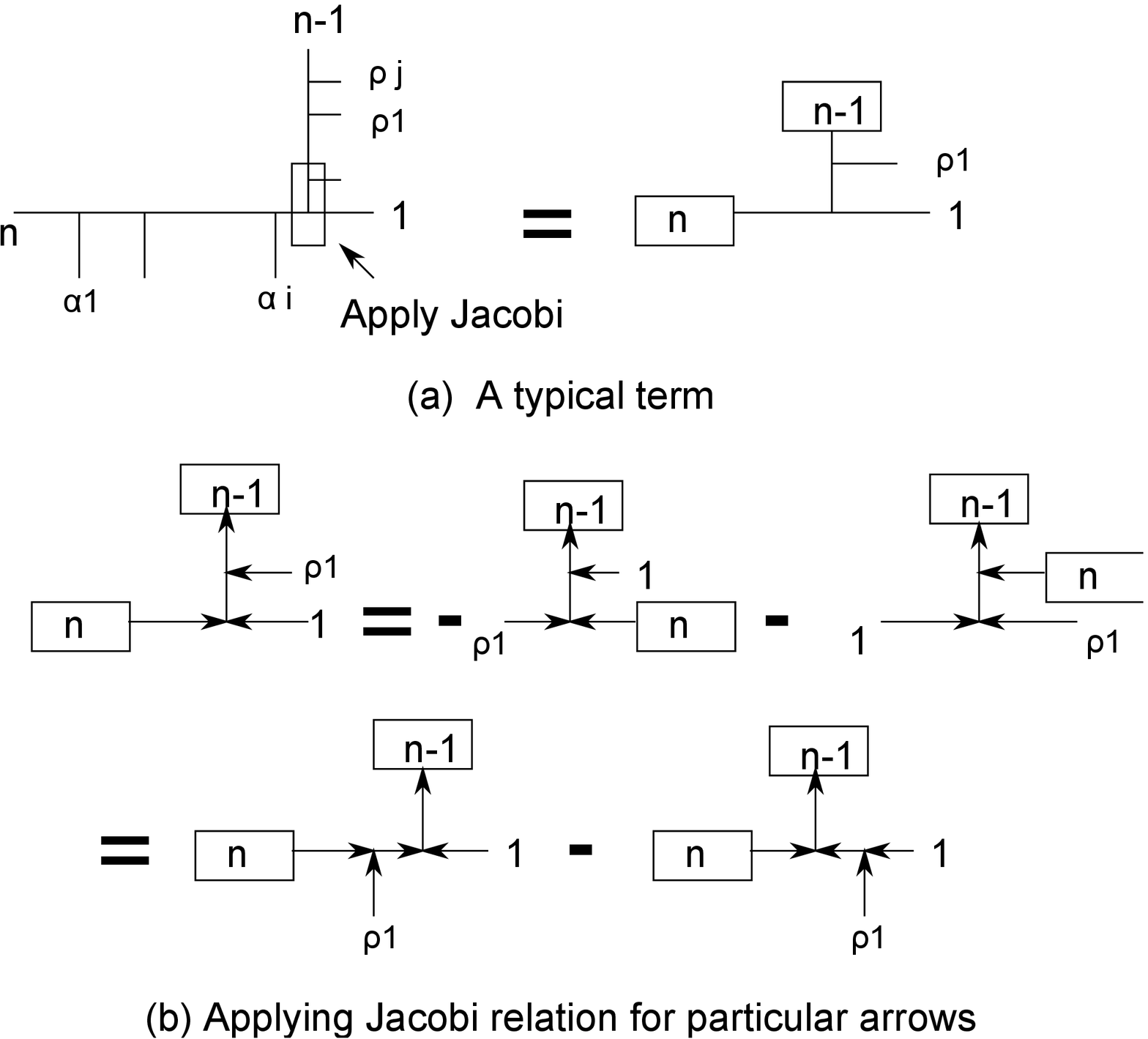,width=13cm}\caption{(a)
Diagrammatic representation of a typical term
$n_{n\alpha_1\dots\alpha_i1\rho_{1}\dots\rho_{j}(n-1)}$ and its
simplified schematic.  On the left hand side we
 use a square to highlight the place where Jacobi identity
is subsequently applied in the graphs below. 
(b)
The manipulation using Jacobi identity on a particular arrow
assignment. For simplicity, we   use an ``$n$-box''   to schematically
represent the chain $\{n,a_1,...,a_i\}$ drawn in (a) and similarly
an ``$(n-1)$-box'' to represent the chain $\{n-1,p_j,...,p_2\}$.  }~~\label{fig-proof}}

Although identities \eref{DDM-relations-1}
 and  \eref{DDM-relations} are well known for experts in the field, it
  nevertheless did not appear in the literature explicitly.
  Since these two identities are very important to the discussion of the relabeling property of
$\tau$-functions constructed by the algorithm outlined in previous sections, 
to be self-contained, we
 would like to review their proofs using the method
 given  in \cite{DelDuca:1999rs}. As we have mentioned,
  the $c_j\eps(q_j)$-part is the
same for all orderings, so we just need to prove that the part in the bracket of
\eref{An-DDM} satisfies above two identities. For this purpose we draw
the diagram representing a typical term  in part
(a) of the Figure \ref{fig-proof} and apply   Jacobi identity to the
part framed by a box. The result is give in  part (b) of Figure
\ref{fig-proof} for a particular arrow configuration. It can be
shown that the same result is true for all other possible arrow
configurations. The result in part (b) tells us that  we can move
the $\rho_1$ attached to the $(n-1)$-th block to two places. At  the
first place $\rho_1$ will be attached to $n$-th block with $+$ sign and
at the second
place $\rho_1$ will be attached to leg $1$ with $-$ sign. Repeating the above
manipulations, we can move down $\rho_2$, $\rho_3$ and finally $\rho_j$. A typical
final configuration will be the ordering $\{n,
\a_1,...,\a_i,\gamma_1,...,\gamma_s,
n-1,\b_r,...,\b_1,1\}$ with sign $(-)^{r+1}$ (the extra
$-$ sign comes from pulling $n-1$ from up to down by antisymmetry), where
sets $\{\gamma_1,...,\gamma_s\}$ and $\{\b_1,...,\b_r\}$ come from the
splitting of original set $\{\rho_1,...,\rho_j\}$ with relative
ordering kept in each subset.
By now, we have proved the identity  \eref{DDM-relations-1}.
Finally we reverse the ordering to get
$\{1,
\b,n-1,\gamma_s^T,\a^T,n\}$
with sign $(-)^{r+1+(n-2)}$. Using $n-3=i+j$ and $j=r+s$ we get the
sign to be $(-)^{i+s}$ (where $T$ means the reversing of ordering).
Thus   identity \eref{DDM-relations} is proved.

Identities \eref{DDM-relations-1} and
\eref{DDM-relations} can also be replaced by the following two forms (where
for later applications we have switched the $1\leftrightarrow n$)
\bea n_{1\alpha_1\dots\alpha_i n\rho_{1}\dots\rho_{j}(n-1)}
=-n_{1 \a_1...\a_i [\rho_1,[\rho_2,...[\rho_{j},(n-1)]...]]n }.
~~\label{DDM-relations-1-1} \eea
and
\begin{equation}
n_{\rho_{1}\dots\rho_{j} 1,\a_1,\dots,\a_i,n}=(-)\,
n_{1[[[\rho_1,\rho_2],...],\rho_j],\a_1,...,\a_i,n}\,,\label{DDM-relations-1-2}
\end{equation}
and $[~,~]$  is the anti-commutative bracket, i.e., $n_{...[a,b]...}=
n_{...ab...}-n_{...ba...}$. Furthermore, if the set
$\rho$ is empty, $[\rho_1,[\rho_2,...[\rho_{j},(n-1)]...]]$ is just
$(n-1)$. Using formula \eref{DDM-relations-1-1} and
\eref{DDM-relations-1-2} we  can move one index to the last
or to the first position.

 There is one remark before we conclude this section. It will be clear soon that
all discussions in this paper are based on the above two identities
\eref{DDM-relations-1} and \eref{DDM-relations}. Although the derivation
has explicitly used a local diagram construction following the approach
in \cite{DelDuca:1999rs}, it seems that the
local diagram construction is not essential since all we need are
Jacobi identities of these BCJ numerators $n_\a$ plus antisymmetry.

\section{ Construction of $\tau$ by BCJ numerators }
\label{sec:const-ntr}
In this section, we present the construction of $\tau$-functions using \eqref{n-tau-relation}, \eref{cyclic-symmetry} and \eqref{KK-relation}.
With $(1,n)$ fixed and  a given ordering $\sigma_2$,...,$\sigma_{n-2}$, the $\tau$ and $n$
are related by \eqref{n-tau-relation}
\bea
n_{1\sigma_2...\sigma_{n-1}n}=\tau_{1[\sigma_2,[...,[\sigma_{n-1},n]...]]},
~~\label{n-tau-relation-2} \eea
where the bracket $[A,B]\equiv AB-BA$. For example, we have
\bean
\tau_{1[2,[3,4]]}=\tau_{1234}-\tau_{1243}-\tau_{1342}+\tau_{1432}.\eean
The expansion at the right handed side of \eref{n-tau-relation-2} includes
cases where $n$ is not at the last index. Thus we should use KK-relation \eqref{KK-relation}
to put $n$ to the last position. After such manipulations we   get
\bea
n_{1\sigma_2...\sigma_{n-1}n}=\Sl_{\sigma'\in S_{n-2}}G_{1,n}(\sigma|\sigma')\tau_{1\sigma'n}.
\label{n-tau-relation3} \eea
To solve $\tau$ by $n$, we need to understand the matrix $G_{1,n}(\sigma|\sigma')$.
The $(n-2)!\times (n-2)!$ matrix can be obtained by the following
steps. First it is easy to see that
\bea \tau_{1[\sigma_2,[...,[\sigma_{n-1},n]...]]}=
\Sl_{\{\sigma\}\in OP(\{\alpha\} \bigcup\{\beta\})}(-1)^{n_{\beta}}\tau_{1\alpha n\beta^T}~~~\label{tau-exp}
\eea
where $n_\beta$ is the number of elements of the set $\b$  and the
sum has the same meaning as the sum in \eref{DDM-relations-1}
 and  \eref{DDM-relations}, i.e., it is over all possible splittings of the ordered set
$\{\sigma_2,...,\sigma_{n-1}\}$ in two subsets $\a$ and $\b$ (empty
sets are allowed) such that inside each subset the relative ordering
defined by the  set $\sigma$ is kept. For example, the set
$\sigma=\{234\}$ has the following eight splittings
{\small\bean (\a,\b)=( \{234\}, \emptyset)/(\{23\},\{4\})/(\{24\},\{3\})/
(\{34\},\{2\})/(\emptyset,\{234\})/(\{4\},\{23\})/(\{3\},\{24\})/
(\{2\},\{34\}) \eean}
Secondly using the imposed KK relation on $\tau$ \eref{KK-relation}, any
$(-1)^{n_{\beta}}\tau_{1\alpha n\beta^T}$ can be expressed as
$\Sl_{\rho\in OP(\{\alpha\}\bigcup\{\beta\})}\tau_{1\rho n}$.
Combining these two steps we arrive at
\bea
\tau_{1[\sigma_2,[...,[\sigma_{n-1},n]...]]}=\Sl_{\{\sigma\}\in OP(\{\alpha\} \bigcup\{\beta\})}
\Sl_{\{\rho\}\in OP(\{\alpha\}\bigcup\{\beta\})}\tau_{1\rho n}.
\eea
From this formula we can see  that the matrix element $G_{1,n}(\sigma|\sigma')$ is
the number of splittings of $\sigma$ into two subsets $\a,\b$, such that
the ordering $\sigma'$ can be obtained by  recombining two subsets $\a,\b$ arbitrarily with
relative ordering kept inside each subset.

Let us give a few examples of $G_{1,n}(\sigma|\sigma')$ with
$\sigma=\{234\}$. For $\sigma'=\{234\}$, there are eight splittings
of $\sigma$, which can be used to recombine to get $\sigma'$, so
$G_{1,5}(\{234\}|\{234\})=8$.
 For
$\sigma'=\{243\}$ there are four splittings $(\{23\},\{4\})$,
$(\{24\},\{3\})$ ,  $(\{4\},\{23\})$, $(\{3\},\{24\})$ available, so $G_{1,5}(\{234\}|\{243\})=4$. For
$\sigma'=\{423\}$ there are only two splittings $(\{23\},\{4\})$,
  $(\{4\},\{23\})$ available, so $G_{1,5}(\{234\}|\{423\})=2$.
  For $\sigma'=\{432\}$, there is no any splitting, so $G_{1,5}(\{234\}|\{432\})=0$. In fact, one can
show that for three numbers $i,j,k$, if their ordering inside the set
$\sigma$ is $i>j>k$ and their ordering inside the set $\sigma'$ is
$i<j<k$, the $G_{1,n}(\sigma|\sigma')=0$. The reason is the following.
To reproduce the right ordering inside $\sigma'$, when we split
$\sigma$ into two subsets, elements $i,j$ must belong to different subsets.
 Similar distributions must  hold for pairs $i,k$ and $j,k$, but these three conditions
 can not be satisfied simultaneously.
 From this general argument, we can see that
the majority elements of matrix $G_{1,n}(\sigma|\sigma')$ are zero.

Now let us discuss some general properties of matrix
$G_{1,n}(\sigma|\sigma')$:
\begin{itemize}

\item (1) First the value can be written as
\bea
G_{1,n}(\sigma|\sigma')=\Sl_{ \{\sigma'\}\in OP( \{\alpha'\}\{\beta'\})}
\Sl_{ \{\sigma\}\in OP(\{\alpha\}\bigcup \{\beta\})}\delta(\{\alpha\},\{\beta\}|\{\alpha'\},\{\beta'\}),~~\label{G-delta-1}
\eea
where these two sums  are over all the possible ordered sets $\alpha$, $\beta$ such that $\{\sigma\}\in OP(\{\alpha\}\bigcup \{\beta\})$ and
all the possible ordered sets $\{\alpha'\}$, $\{\beta'\}$ such that $\{\sigma'\}\in OP( \{\alpha'\}\{\beta'\}$.
In other word, we sum over all the possible splittings of $\sigma$ and $\sigma'$ into  two ordered subsets respectively.
This means   elements of matrix $G_{1,n}(\sigma|\sigma')$ can be calculated by another way. First we find all $2^{n_\sigma}$ splittings of the set $\sigma$ to
two subsets $(\a,\b)$ with relative ordering kept and all splittings of the set $\sigma'$ into two subsets $(\a',\b')$ with relative
ordering kept. Then we find all subsets such that $(\a,\b)=(\a',\b')$. The total number is $G_{1,n}(\sigma|\sigma')$.

\item  (2) From the above property \eref{G-delta-1}, it is easy to see that $G$ is symmetric
\bea
G_{1,n}(\sigma|\sigma')=G_{1,n}(\sigma'|\sigma).~~~\label{G-symmetric}
\eea

\item (3) Some special elements can be obtained. When $\sigma=\sigma'$, all possible splittings of $\sigma\in S_{n-2}$ should be counted,
i.e., $G_{1,n}(\sigma|\sigma)=2^{n-2}$. If $\sigma,\sigma'$ are
different only by one permutation of two adjacent numbers, then
$G_{1,n}(\sigma|\sigma')=2^{n-3}$. If $\sigma, \sigma'$ have the
following orderings: $\sigma=(...,i_1,i_2,...,j_1,j_2,...)$ and
$\sigma'=(...,i_2,i_1,...,j_2,j_1,...)$, we have
$G_{1,n}(\sigma|\sigma')=2^{n-4}$. Similar pattern holds for more
interchanges of adjacent pairs.

\item (4) The fourth observation is that
\bea
G_{1,n}(\sigma|\sigma')=G_{1,n}(P(\sigma)|P(\sigma'))~~~~~\label{G-permu}\eea
where $P$ is any permutation of $(n-2)$ elements. The reason is that
\eref{G-delta-1} cares only  relative orderings between
sets $\sigma, \sigma'$, so it does not matter if a  element is
called $x$ or $y$.

\end{itemize}

Having discussed the matrix $G_{1,n}(\sigma|\sigma')$,  we can solve
$\tau_{1,\sigma,n}$ by BCJ numerators using
\eqref{n-tau-relation3} and obtain
\bea \tau_{1\sigma' n}=\frac{\det\mathbb{G'}}{\det\mathbb{G}},~~~~\label{tau-solu} \eea
where $\mathbb{G}=G_{1,n}(\sigma|\sigma')$ is the $(n-2)!\times(n-2)!$ matrix,
and $\mathbb{G'}$ is obtained by
replacing the column $\sigma'$ of $\mathbb{G}$ by the column of
BCJ numerators  $n_{1\sigma n}$. Now we have provided a general
construction of $(n-2)!$ $\tau$'s with $1$, $n$ fixed at the two ends \eref{tau-solu}.
Other $(n!-(n-2)!)$ $\tau$'s can be obtained by KK relations and cyclic symmetry
 (see \eref{cyclic-symmetry} and \eqref{KK-relation}).

The above construction for $\tau$'s
 is complete and there is no any ambiguity.
The only subtle point is that in the construction, two special elements have been fixed,
for example $1$ and $n$. Then it is natural to ask what is the relation
between two solutions obtained by fixing two different pairs  $(i_1, j_1)$ and
$(i_2, j_2)$?  There are two possibilities. The first possibility is that these two solutions do not have
any relation. Another possibility is that these two solutions will relate to each other by
some manipulations. As we will show in later sections, there is a {\bf natural relabeling property}
for solutions obtained by above construction. This property tells us that if we have
an expression for just one $\tau$-function, expressions for all  other $\tau$'s
can be obtained by relabeling. Furthermore,
all solutions coming from different fixed pairs will give the same answer.

\section{Examples}
\label{sec:example}
In this section, we will use several examples to explicitly demonstrate
the algorithm for  $\tau$-functions and check
that the solution satisfies the  natural relabeling property.


\subsection{Three-point case}
In this case, with $1,3$ fixed we have $G_{1,3}(\{2\}|\{2\})=2$,
\bea
n_{123}=2\tau_{123}\Longrightarrow \tau_{123}=\frac{1}{2}n_{123}.~~~\label{tau123}
\eea
Other five $\tau$'s can be obtained from $\tau_{123}$ by cyclic symmetry and
KK-relations
\bea {\rm Cyclic}: & ~~ & \tau_{312}=\tau_{231}=\tau_{123}\nn {\rm
KK-rel}: & ~~&
\tau_{213}=\tau_{321}=\tau_{132}=-\tau_{123}~~\label{KK-cyc} \eea

To check the relabeling, noticing that if we exchange $3\leftrightarrow 2$
in \eref{tau123}, we   get
\bea \W\tau_{132}={1\over 2} n_{132}\eea
Because for  three-point case, BCJ numerator $n$ is cyclic symmetric and anti-symmetric under
exchanging of pair, i.e.,
$n_{123}=-n_{132}$,  we see that $\W\tau_{132}$ is equal to
$\tau_{132}$ in \eref{KK-cyc}, i.e., $\tau_{132}$ can be obtained from
 $\tau_{123}$ by relabeling indices. It is easy to check the
relabeling property for other $\tau$'s.

\subsection{Four-point case}

For four-point case, we have $4!=24$ $\tau$'s. Now let us use our
algorithm to determine all of them and check the relabeling
property:

~\\{\bf Solving the $\tau$'s in KK-basis:} we have $(4-2)!=2$ equations and
$(4-2)!=2$ independent $\tau$'s. With ordering $(\{23\},\{32\})$ the
matrix is
\bea G_{1,4}=\left( \begin{array}{cc} 4 & 2 \\ 2 & 4
\end{array}\right)~~~~\label{G-4}\eea
From this we can solve
\bea \tau_{1234}= {  \left| \begin{array}{cc} n_{1234} & 2 \\
n_{1324} & 4
\end{array}\right|\over
 \left| \begin{array}{cc} 4 & 2 \\ 2 & 4
\end{array}\right|}=\frac{1}{3}n_{1234}-\frac{1}{6}n_{1324} ~~~\label{4p-1234}\eea
and
\bea \tau_{1324}= {  \left| \begin{array}{cc} 4 & n_{1234}  \\ 2 &
n_{1324}
\end{array}\right|\over
 \left| \begin{array}{cc} 4 & 2 \\ 2 & 4
\end{array}\right|}=-\frac{1}{6}n_{1234}+\frac{1}{3}n_{1324} ~~~\label{4p-1324}\eea

~\\{\bf Find remaining $\tau$'s}: Having  $\tau_{1234}$ and $\tau_{1324}$, we can construct
other $22$ $\tau$'s. Using   KK-relation, we can
obtain the following four $\tau$ with $1$ fixed:
\bea \tau_{1243}& \equiv &  -\tau_{1234}-\tau_{1324}=-{1\over 6}
n_{1234}-{1\over 6} n_{1324} \nn
\tau_{1342}& \equiv &  -\tau_{1234}-\tau_{1324}=-{1\over 6}
n_{1234}-{1\over 6} n_{1324} \nn
\tau_{1423}& \equiv &
+\tau_{1324}=-\frac{1}{6}n_{1234}+\frac{1}{3}n_{1324} \nn
\tau_{1432}& \equiv &
+\tau_{1234}=\frac{1}{3}n_{1234}-\frac{1}{6}n_{1324}
~~~\label{4p-4rem}\eea
Having obtained all $\tau_{1\sigma}$, the remaining $18$ $\tau$'s
are obtained by {  imposed cyclic symmetry}. For example,
we have
\bea
\tau_{1234}=\tau_{4123}=\tau_{3412}=\tau_{2341}=\frac{1}{3}n_{1234}-\frac{1}{6}n_{1324}
~~~\label{tau4-4123}\eea

~\\{\bf Relabeling property:} Using these explicit result, we can check the relabeling
property. First it is easy to see that  expression \eref{4p-1324}
can be obtained from \eref{4p-1234} by replacing $2\to 3, 3\to 2$.
In principle, this fact does not need to hold. However, by symmetric
property and property \eref{G-permu}  of matrix $G$, it is natural to get
  $\tau_{1 P(\sigma)n}=
P(\tau_{1\sigma n})$.

Now we check the relabeling property for $\tau_{1\sigma}$. From
solution $\tau_{1234}$ by index relabeling $(3,4)\to (4,3)$
 we can write down
\bea \W\tau_{1243}= \frac{1}{3}n_{1243}-\frac{1}{6}n_{1423}\eea
To check if  $\W \tau_{1243}$ is equal to $\tau_{1243}$ obtained in
\eref{4p-4rem}, we  use   relation
\eref{DDM-relations-1} to get
\bea n_{1243}=-n_{1234},~~~n_{1423}=-n_{1234}+n_{1324}\eea
Putting it back, it is easy to check $\W \tau_{1243}=\tau_{1243}$. In other
words, two expressions, i.e., the one obtained by our imposed
KK-relation and the one obtained by index relabeling from known
solution $\tau_{1\sigma 4}$, are the same!

Next, we check the relabeling property for $\tau$'s where $1$ is not the
first index. By relabeling $(1,2,3,4)\to
(4,1,2,3)$ from $\tau_{1234}$ we have
\bea \W
\tau_{4123}=\frac{1}{3}n_{4123}-\frac{1}{6}n_{4213}.~~~\label{4p-4123}
\eea
Using   relation
\eref{DDM-relations}, this expands into
\bea n_{4123}=n_{1234}-n_{13(24)},~~~ n_{4213}=-n_{1324}, \eea
and we do find $\W\tau_{4123}$ is equal to the $\tau_{4123}$ written down in \eref{tau4-4123}.

Although we have presented  only three examples,
using the Mathematica we have  checked that indeed the result obtained from our algorithm (including
using KK and cyclic relations) does agree with the one obtained from relabeling of single
$\tau_{1234}$ expression.

\subsection{Five-point case}

Having seen the above four-point example, we will not present too much
detail for cases with five, six and seven points. The rest can be easily
recovered via the same algorithm.

~\\{\bf Solutions from our algorithm:} First we can solve $\tau$'s in KK-basis using
the following equation
\bea
\left(
  \begin{array}{c}
    n_{12345} \\
    n_{12435} \\
    n_{13245} \\
    n_{14235} \\
    n_{13425} \\
    n_{14325} \\
  \end{array}
\right)&=&\left(
          \begin{array}{cccccc}
           2^3 & 2^2 & 2^2 & 2^1 & 2^1 & 0 \\
           2^2 & 2^3 & 2^1 & 2^2 & 0   & 2^1 \\
           2^2 & 2^1 & 2^3 & 0   & 2^2 & 2^1 \\
           2^1 & 2^2 & 0   & 2^3 & 2^1 & 2^2 \\
           2^1 &0    & 2^2 & 2^1 & 2^3 & 2^2 \\
            0  & 2^1 & 2^1 & 2^2 & 2^2 & 2^3 \\
          \end{array}
        \right)\left(
                 \begin{array}{c}
                    \tau_{12345} \\
                   \tau_{12435} \\
                   \tau_{13245} \\
                   \tau_{14235} \\
                   \tau_{13425} \\
                   \tau_{14325} \\
                 \end{array}
               \right).~~~~\label{eq:5-pt-g}
\eea
and obtain an expression of $\tau_{12345}$  as
\bea
\tau_{12345}=\frac{1}{4}n_{12345}-\frac{1}{10}n_{12435}-
\frac{1}{20}n_{14235}-\frac{1}{10}n_{13245}-\frac{1}{20}n_{13425}+\frac{1}{10}n_{14325}.~~\label{tau5p-12345}
\eea
Other five $\tau$'s can be obtained by relabeling $2,3,4$ from \eref{tau5p-12345}. Having these six  $\tau$ solutions
for KK-basis, we can use the KK-relations and cyclic relations to find other $114$ $\tau$'s.

~\\{\bf Relabeling property:} Now we need to check if the above result has the relabeling property. Although we have used Mathematica to check that
all $120$ $\tau$'s are related to each other by relabeling property, here we will give only a few examples.
The first example is\footnote{It is worth  noticing that the expression of $\tau_{12534}$ is simpler
than \eref{tau5p-12345} because it has only $5$ terms and simpler coefficients. It will
be interesting to investigate if it is
 general.  }
\bea \tau_{12534}& = & -{1\over 20} n_{12345}+{1\over 10} n_{12435}-{1\over 20} n_{13425}
+{1\over 20} n_{14235}+{1\over 10} n_{14325}~~~\label{tau-5p-12534}\eea
which is obtained from KK-relation. Let us relabel the solution \eref{tau5p-12345} with $3\to 5, 4\to 3,
5\to 4$. After plugging
\bean n_{12534} & = & -n_{12345}+n_{12435} \nn
n_{12354} & = & -n_{12345} \nn
n_{13254} & = & -n_{13245} \nn
n_{15234} & = & -n_{12345}+n_{12435}+n_{13425}-n_{14325}\nn
n_{15324} & = & n_{12435}-n_{13245}+n_{13425}-n_{14235}\nn
n_{13524} & = & -n_{13245}+n_{13425}\eean
it can be checked that the relabeling \eref{tau5p-12345} does reproduce   \eref{tau-5p-12534}.

The second example is $\tau_{51234}\equiv \tau_{12345}$ by our definition. The relabeling of \eref{tau5p-12345}
gives
\bea
\W\tau_{51234}=\frac{1}{4}n_{51234}-\frac{1}{10}n_{51324}-\frac{1}{20}n_{53124}-\frac{1}{10}n_{52134}
-\frac{1}{20}n_{52314}+\frac{1}{10}n_{53214}.
\eea
Using the result
\bea
n_{51234}&=&n_{12345}-n_{12435}-n_{13425}+n_{14325},\nn
n_{51324}&=&n_{13245}-n_{13425}-n_{12435}+n_{14235},\nn
n_{53124}&=&-n_{12435}+n_{14235},\nn
n_{52134}&=&-n_{13425}+n_{14325},\nn
n_{52314}&=&n_{14325},\nn
n_{53214}&=&n_{14235}.
\eea
it is easy to see that $\tau_{51234}=\W\tau_{51234}$, i.e., the relabeling property is satisfied.

\subsection{Six-point case}

For six points, using the algorithm we can solve $24$ $\tau$'s in   KK-basis. Since they
are related to each other by relabeling, we simply write down   one solution $\tau_{123456}$,
\bea
\tau_{123456}&=&\frac{1}{630}\Bigl[ 126n_{123456}-44n_{123546}-44n_{124356}-19n_{124536}-19n_{125346}+36n_{125436}\nn
&&-44n_{132456}+16n_{132546}-19n_{134256}-19n_{134526}+n_{135246}+11n_{135426}\nn
&&-19n_{142356}+n_{142536}+36n_{143256}+11n_{143526}-9n_{145236}+16n_{145326}\nn
&&-19n_{152346}+11n_{152436}+11n_{153246}+16n_{153426}+16n_{154236}-44n_{154326}
\Bigr].~~\label{tau-123456}
\eea

Using   KK-basis of $\tau$ we can get the remaining $696$ $\tau$'s.
Now let us check the relabeling property. We have used the
Mathematica to check that all $720$ $\tau$'s are related to each
other by the relabeling property. Here
 we give only one example $\tau_{612345}\equiv \tau_{123456}$ by
cyclic symmetry. The relabeling from $\tau_{123456}$ gives
\bea
\W \tau_{612345}&=&\frac{1}{630}\Bigl[126n_{612345}-44n_{612435}-44n_{613245}-
19n_{613425}-19n_{614235}+36n_{614325}\nn
&&-44n_{621345}+16n_{621435}-19n_{623145}-19n_{623415}+n_{624135}+11n_{624315}\nn
&&-19n_{631245}+n_{631425}+36n_{632145}+11n_{632415}-9n_{634125}+16n_{634215}\nn
&&-19n_{641235}+11n_{641325}+11n_{642135}+16n_{642315}+16n_{643125}-44n_{643215}
\Bigr].
\eea
Using the result
\bea
n_{61ijk5}&=&n_{1ijk56} - n_{1jk5i6} - n_{1ik5j6} - n_{1ij5k6} +n_{1k5ji6} + n_{1i5kj6} + n_{1j5ki6} - n_{15kji6}\nn
n_{6i1jk5}&=&-n_{1jk5i6} + n_{1j5ki6} + n_{1k5ji6} -n_{15kji6}\nn
n_{6ij1k5}&=&n_{1k5ji6} - n_{15kji6}\nn
n_{6ijk15}&=&-n_{15kji6},
\eea
and indeed we   see that $\W\tau_{612345}=\tau_{612345}$.

\subsection{Seven-point case}
At seven-points, the expression for $\tau_{1234567}$ with $1,7$ fixed is
\bea
\tau_{1234567}&={1\over 24192}&
\Bigl(4032 n_{1234567} -1284 n_{1234657} -1284 n_{1235467} -513 n_{1235647} -513 n_{1236457} +940 n_{1236547}\nn
&& -1284 n_{1243567} +393 n_{1243657}-513 n_{1245367} -438 n_{1245637} +27 n_{1246357} +259 n_{1246537}\nn
&& -513 n_{1253467} +27 n_{1253647} +940 n_{1254367} +259 n_{1254637} -213 n_{1256347} +337 n_{1256437}\nn
&& -438 n_{1263457} +259 n_{1263547} +259 n_{1264357} +344 n_{1264537} +337 n_{1265347} -940 n_{1265437}\nn
&& -1284 n_{1324567} +421 n_{1324657} +393 n_{1325467} +205 n_{1325647} +205 n_{1326457} -337 n_{1326547}\nn
&& -513 n_{1342567} +205 n_{1342657} -438 n_{1345267} -513 n_{1345627} +32 n_{1346257} +205 n_{1346527}\nn
&& +27 n_{1352467} -56 n_{1352647} +259 n_{1354267} +184 n_{1354627} +23 n_{1356247} +157 n_{1356427}\nn
&& +32 n_{1362457} -6 n_{1362547} -51 n_{1364257} +143 n_{1364527} +6 n_{1365247} -259 n_{1365427}\nn
&& -513 n_{1423567} +205 n_{1423657} +27 n_{1425367} +32 n_{1425637} -56 n_{1426357} -6 n_{1426537}\nn
&& +940 n_{1432567} -337 n_{1432657} +259 n_{1435267} +205 n_{1435627} -6 n_{1436257} -148 n_{1436527}\nn
&& -213 n_{1452367} +23 n_{1452637} +337 n_{1453267} +157 n_{1453627} -213 n_{1456237} +337 n_{1456327}\nn
&& +23 n_{1462357} -53 n_{1462537} +6 n_{1463257} +11 n_{1463527} +148 n_{1465237} -205 n_{1465327}\nn
&& -438 n_{1523467} +32 n_{1523647} +259 n_{1524367} -51 n_{1524637} +23 n_{1526347} +6 n_{1526437}\nn
&& +259 n_{1532467} -6 n_{1532647} +344 n_{1534267} +143 n_{1534627} -53 n_{1536247} +11 n_{1536427}\nn
&& +337 n_{1542367} +6 n_{1542637} -940 n_{1543267} -259 n_{1543627} +148 n_{1546237} -205 n_{1546327}\nn
&& -213 n_{1562347} +148 n_{1562437} +148 n_{1563247} +107 n_{1563427} +107 n_{1564237} -393 n_{1564327}\nn
&& -513 n_{1623457} +205 n_{1623547} +184 n_{1624357} +143 n_{1624537} +157 n_{1625347} -259 n_{1625437}\nn
&& +205 n_{1632457} -148 n_{1632547} +143 n_{1634257} +344 n_{1634527} +11 n_{1635247} -184 n_{1635427}\nn
&& +157 n_{1642357} +11 n_{1642537} -259 n_{1643257} -184 n_{1643527} +107 n_{1645237} -421 n_{1645327}\nn
&& +337 n_{1652347} -205 n_{1652437} -205 n_{1653247} -421 n_{1653427} -393 n_{1654237} +1284 n_{1654327}\Bigr).\nn \eea
By relabeling  the above expression $(1,2,3,4,5,6,7)\rightarrow (7,1,2,3,4,5,6)$, we get $\W\tau_{7123456}$.
To show that it is equal to  $\tau_{7123456}=\tau_{1234567}$ from cyclic symmetry,
we just need to  use the following expressions
\bea
n_{71ijkl6}&=&n_{1ijkl67} - n_{1ijk6l7} -n_{1ijl6k7} + n_{1ij6lk7} -n_{1ikl6j7} + n_{1ik6lj7}+n_{1il6kj7} - n_{1i6lkj7}\nn
 &-&n_{1jkl6i7} + n_{1jk6li7} +n_{1jl6ki7} - n_{1j6lki7} +n_{1kl6ji7} - n_{1k6lji7} -n_{1l6kji7} + n_{16lkji7}\nn
n_{7i1jkl6}&=&-n_{1jkl6i7} + n_{1jk6li7} +n_{1jl6ki7} - n_{1j6lki7} +n_{1kl6ji7} - n_{1k6lji7} -n_{1l6kji7} + n_{16lkji7}\nn %
n_{7ij1kl6}&=&n_{1kl6ji7} - n_{1k6lji7} -n_{1l6kji7} + n_{16lkji7}\nn %
n_{7ijk1l6}&=&-n_{1l6kji7} + n_{16lkji7}\nn
n_{7ijkl16}&=&n_{16lkji7} \eea

\section{Relabeling property}
\label{sec:rel-propty} In the previous section, we have demonstrated
the consistency between our algorithm and the relabeling property.
In this section, we will give a general understanding of this
property. Before doing so, we need to address a technical issue
concerning the definition of matrix $G$. Note that generically
relabeling is a permutation of $n$ elements while the definition of
$G$ involves only permutation of $(n-2)$ elements since two of them
have been fixed. To relate matrix $G$ with different fixed pairs, we
enlarge the definition of matrix $G_{1,n}(\sigma|\sigma')$ to
$G_{1,n}(\sigma|\sigma')\equiv G(1,\sigma,n|1,\sigma',n)$ and
require that when we split the set $\{1,\sigma,n\}$ to two subsets,
the first element $1$ must be at the first subset and the last
element $n$ must be at the second subset. It is easy to see that the
new definition is equivalent to the old one. More importantly, the
relabeling property \eref{G-permu} of $(n-2)$ elements can be
enlarged to incorporate the relabeling of $n$ elements, i.e, we will
have that
\bea G(\sigma|\sigma')=G(P(\sigma)|P(\sigma')),~~~~P\in S_n~,~~~\label{G-permu-1}\eea
 where now $\sigma,\sigma'$ are lists of $n$ elements.

Having enlarged the definition of matrix $G$, we can start our discussions. The structure of
this section is the following. First we will set up a general framework for discussions. Then we will
discuss how the relabeling property can be used to solve $\tau$-function.

\subsection{The proof of relabeling property}

We note that   the relabeling property can be proven without knowing the solution explicitly.
Since by our algorithm  three equations
\eqref{n-tau-relation}, \eref{cyclic-symmetry} and \eqref{KK-relation} fix the solution
uniquely, we can prove the   property by showing that
it is a property possessed by these three equations.
The imposed cyclic symmetry \eref{cyclic-symmetry} and KK-relations
 \eqref{KK-relation} have this property obviously, thus the key is to show that
it is a property of
 equation \eqref{n-tau-relation}
 as well.

Consider the relabeling $x_s\to z_s$ with $s=1,...,n$.
Under the relabeling, an ordering of $n$ elements becomes another
ordering of $n$ elements,   $X_i\to Z_i$ with $i=1,...,(n-2)!$
running through the whole KK-basis. Thus equation \eref{n-tau-relation3}
becomes
\bea \sum_{j=1}^{(n-2)!} G_{X_i X_j} \tau_{X_j}=n_{X_i} \Longrightarrow \sum_{j=1}^{(n-2)!} G_{Z_i Z_j} \tau_{Z_j}=n_{Z_i}
~~~~\label{n-tau-rel-1}\eea
where the sum  is  over KK-basis $X_j$.  However, by   cyclic and KK-relations, we have
\bea \tau_{Z_i}=\sum_{j=1}^{(n-2)!}Q_{Z_i X_j} \tau_{X_j}~~~\label{tau-relabel}\eea
and by \eref{DDM-relations} and \eref{DDM-relations-1} we have
\bea  n_{Z_i}= \sum_{j=1}^{(n-2)!}P_{Z_i X_j} n_{X_j}~~~\label{n-relabel}\eea
thus
\bea  G_{Z_i Z_j} \tau_{Z_j}=n_{Z_i} \Longrightarrow  G_{Z_i Z_j}
Q_{Z_j X_k} \tau_{X_k}=P_{Z_i X_l} n_{X_l}\eea
where to make the notation simpler, we have used Einstein summation convention.

To see that solution from our algorithm   have the relabeling property, we just need to show
in addition,
\bea  [P_{Z_i X_l}]^{-1} G_{Z_i Z_j} Q_{Z_j X_k}=G_{X_l
X_k}~~~\label{G-con-rel}\eea

Now let us demonstrate the use of \eref{G-con-rel} by several examples. First let us consider
the four-point case where the matrix $G$ is given by \eref{G-4}.
For relabeling  given by permutation $P(3,4)$, we have
$X_1=(1,2,3,4)\to Z_1= (1,2,4,3)$ and $X_2=(1,3,2,4)\to
Z_2=(1,4,2,3)$. Under this permutation we   find
\bea \left( \begin{array}{c} \tau_{Z_1} \\
\tau_{Z_2}\end{array}\right)=\left(
\begin{array}{cc} -1 & -1 \\ 0 & 1 \end{array}\right)
\left( \begin{array}{c} \tau_{X_1} \\
\tau_{X_2}\end{array}\right)\eea
and
\bea \left( \begin{array}{c} n_{Z_1} \\
n_{Z_2}\end{array}\right)=\left(
\begin{array}{cc} -1 & 0 \\ -1 & 1 \end{array}\right)
\left( \begin{array}{c} n_{X_1} \\
n_{X_2}\end{array}\right)\eea
Thus it is easy to check that
\bea \left( \begin{array}{cc} 4 & 2 \\ 2 & 4
\end{array}\right)= \left(
\begin{array}{cc} -1 & 0 \\ -1 & 1 \end{array}\right)^{-1}
\left( \begin{array}{cc} 4 & 2 \\ 2 & 4
\end{array}\right)\left(
\begin{array}{cc} -1 & -1 \\ 0 & 1 \end{array}\right)\eea
%

As a second example, we
consider the permutation $P(1,2)$ at five-point,
where the $G$ matrix is given by (\ref{eq:5-pt-g}). For this
permutation, we find the matrix representation of $P$ through the
 following manipulation
\bea
\left(\begin{array}{c}
n_{12345}\\
n_{12435}\\
n_{13245}\\
n_{13425}\\
n_{14235}\\
n_{14325}
\end{array}\right)\rightarrow\left(\begin{array}{c}
n_{21345}\\
n_{21435}\\
n_{23145}\\
n_{23415}\\
n_{24135}\\
n_{24315}
\end{array}\right)=(-)\left(\begin{array}{c}
n_{12345}\\
n_{12435}\\
n_{1[2,3]45}\\
n_{1[[2,3],4]5}\\
n_{1[2,4]35}\\
n_{1[[2,4],3]5}
\end{array}\right)=\left(\begin{array}{cccccc}
-1 & 0 & 0 & 0 & 0 & 0\\
0 & -1 & 0 & 0 & 0 & 0\\
-1 & 0 & 1 & 0 & 0 & 0\\
-1 & 0 & 1 & 0 & 1 & -1\\
0 & -1 & 0 & 0 & 1 & 0\\
0 & -1 & 1 & -1 & 1 & 0
\end{array}\right)\left(\begin{array}{c}
n_{12345}\\
n_{12435}\\
n_{13245}\\
n_{13425}\\
n_{14235}\\
n_{14325}
\end{array}\right).
~~\label{eq:n5-q}
\eea
Similar manipulation gives the matrix of $Q$ as
\begin{equation}
Q_{Z_{i}X_{j}}=\left(\begin{array}{cccccc}
-1 & 0 & -1 & -1 & 0 & 0\\
0 & -1 & 0 & 0 & -1 & -1\\
0 & 0 & 1 & 1 & 0 & 1\\
0 & 0 & 0 & 0 & 0 & -1\\
0 & 0 & 0 & 1 & 1 & 1\\
0 & 0 & 0 & -1 & 0 & 0
\end{array}\right).
\end{equation}
It is straightforward to see that
$[P_{Z_{i}X_{l}}]^{-1}G_{Z_{i}Z_{j}}Q_{Z_{j}X_{k}}=G_{X_{l}X_{k}}$
is indeed satisfied.

Having shown two examples, now we discuss how we could give a general proof.
To do so, we need the following observation. Assuming the relabeling from $X_i\to Z_i$
can be broken into two steps,
$X_i\to Y_i$ and $Y_i \to Z_i$, it is easy to see that
\bea  \tau_{Z_i}=Q_{Z_i Y_t} \tau_{Y_t}=Q_{Z_i Y_t} Q_{Y_t X_j}
\tau_{X_j},~~~n_{Z_i}= P_{Z_i Y_t } n_{Y_t }=P_{Z_i Y_t } P_{Y_t X_j}
n_{X_j},~~~\label{XY-steps-1}\eea
and the condition \eref{G-con-rel} becomes
\bea [P_{Z_i Y_{t^{'}}} P_{Y_{t^{'}} X_l}]^{-1}G_{Z_i Z_j}Q_{Z_j Y_t} Q_{Y_t
X_k}= G_{X_l X_k}~~~\label{XY-steps-2}~.\eea
If the relabeling property holds for both
steps $X_i\to Y_i$ and $Y_i \to Z_i$, i.e.
\bea  [P_{Y_{t^{'}} X_l}]^{-1} G_{Y_{t^{'}} Y_t} Q_{Y_t X_k}=G_{X_l X_k}\eea
and
\bea  [P_{Z_i Y_{t^{'}}}]^{-1} G_{Z_i Z_j} Q_{Z_j Y_t}=G_{Y_{t^{'}} Y_t}\eea
then \eref{XY-steps-2}   also holds. This observation reflects
 the structure of group, so that if we can show \eref{G-con-rel} is true for
 all
generators of permutation group $S_n$,  it will be true for the whole group.

 For permutation group $S_n$,
there are $(n-1)$ generators.  For our convenience, we   choose
the  following  $(n-2)$
generators\footnote{To avoid confusion of matrix $P$ and the generators of permutation group,
we will use ${\cal P}$ for later.}
\bea
{\cal P}_n=(1,n),~~~{\cal P}_i=(i,i+1),~~~~i=2,3,...,n-2;~~~\label{Sn-gen}
\eea
plus any one permutation of the form ${\cal P}_{1i}=(1,i)$ or
${\cal P}_{ni}=(n,i)$ with $i=2,3,...,n-1$. For permutations ${\cal P}_i$ with
$i=2,...,n-2$, they are  permutations among KK-basis with $(1,n)$
fixed and it is easy to see that corresponding matrixes $P,Q$ satisfy
 $P=Q$ and $P^{-1}=P^T$. By the
general property \eref{G-permu}, $P^{-1} G P=
G({\cal P}(\sigma)|{\cal P}(\sigma'))= G(\sigma|\sigma')$, i.e., the condition
\eref{G-con-rel} is satisfied.

For permutation ${\cal P}_n$, since $2,3...,n-2$ are invariant, we have
\bea \tau_{n\sigma 1}=\tau_{1n\sigma}=(-)^{n-2}\tau_{1\sigma^T
n},~~~~n_{n\sigma 1}=(-)^{n-2} n_{1\sigma^T n}\eea
and especially
\bea G(1\sigma^T n|1\gamma^T n)=G(1\sigma n|1\gamma n)~,\eea
thus the relabeling property for permutation ${\cal P}_n$ is proved. To finish the proof,
we need to check that the last permutation ${\cal P}_{1i}$ or ${\cal P}_{ni}$
satisfies the relabeling property \eref{G-con-rel}. Since the proof is very complicated, we
leave it to   Appendix.

\subsection{An application}

Having shown that solution from our algorithm has the  relabeling property, it is natural
to ask if we can use the relabeling property to fix the solution. In this subsection, we will
show that it can be done. There are two different approaches and we   discuss
them one by one.

For the first approach, we need to use the relabeling property and only one equation
of the form \eref{n-tau-relation}.  To demonstrate the idea, let us start with four-point example.
First we expand $\tau_{1234}$ into
the KK-basis $n_{1\sigma n}$ as
\bea \tau_{1234}=\a n_{1234}+\b n_{1324}~~~\label{rel-4-1} \eea
Using the relabeling property other $\tau$'s
in the KK-basis will have
similar expansion. For $n=4$, thing is simple and we have only one
\bea \tau_{1324}& = &\a n_{1324}+\b n_{1234} \eea
To completely fix $\a,\b$ we need to use one relation
\bea n_{1234}&  = &  4 \tau_{1234}+2 \tau_{1324}=(4\a+2\b) n_{1234}+
(4\b+2\a) n_{1324}~~~\label{5-4-1}\eea
Since each BCJ numerator $n_\a$ is independent, we get   two equations
\bea 4\a+2\b=1,~~~~4\b+2\a=0\Longrightarrow \b=-{1\over
6},~~\a={1\over 3}~~~\label{5-4-2} \eea

For general $n$-points, we expand, for example, $\tau_{123...(n-1)n}$
into the KK-basis $n_{1\sigma n}$ of BCJ numerators with $(n-2)!$ unknown variables
$\a_i$. Then we use the relabeling property to express all other $\tau$'s in
KK-basis using the same set of variables $\a_i$. Next we put it back to just one
equation $n_{123...n}=\tau_{1[2,[3,...[n-1,n]]]}$. Identifying both sides   gives $(n-2)!$
linear equations for
coefficients $\a_i$. From these equations we can solve $\a_i$ and determine  expressions
of $\tau$'s.

Now we want to compare the first approach with the algorithm given in section 3.
The algorithm given in section 3 requires calculating a big matrix $G$ and invert it.
However, calculating matrix $G$ by formula \eref{G-delta-1} is not easy
and there are a lot of combinations to bookkeep.
The first approach presented here does not require calculating   $G$. All
information of $G$ is automatically included in the above procedures (see \eref{5-4-1}
and \eref{5-4-2}). In other words, the first approach has bypassed the calculation of
matrix $G$ although solving linear equations of $(n-2)!$ variables can still be a
difficult problem.

The first approach has not used the full potential of relabeling
property. Now we present the second approach. To demonstrate, we use
the five-point case as an example. Expanding $\tau_{12345}$ into the
BCJ basis we have
\bea \tau_{12345}=\a_1 n_{12345}+ \a_2 n_{12435}+\a_3 n_{13245}+\a_4
n_{13425}+ \a_5 n_{14235}+\a_6 n_{14325}~~~\label{rel-5-1}\eea
with six unknown variables.
Using the relabeling property we can write down the expansion of
other five $\tau$'s in   KK-basis using the same six variables $\a_i$.
Up to this step, it is the same as in the first approach. However, we have not
used all generators of permutation group $S_5$, i.e., there
 are still two relabelings ${\cal P}(1,5)$ and ${\cal P}(5,4)$   not
 being used. Using the relabeling property coming from    ${\cal P}(1,5)$, we have on the one hand
\bea \tau_{12345}\to \tau_{52341}& = & \a_1 n_{52341}+ \a_2
n_{52431}+\a_3 n_{53241}+\a_4 n_{53421}+ \a_5 n_{54231}+\a_6
n_{54321}\eea
by relabeling property, but on the other hand
\bea \tau_{12345}\to \tau_{52341}& = & \tau_{15234}=-\tau_{14325}\nn
& = & -(\a_1 n_{14325}+ \a_2 n_{14235}+\a_3 n_{13425}+\a_4
n_{13245}+ \a_5 n_{12435}+\a_6 n_{12345}) \eea
by   cyclic symmetry and  KK-relation of $\tau$.
Comparing these two results using $n_{1\sigma n}=(-)^n n_{n\sigma^T
1}$, we  immediately get the following  equations
\bea \a_1=\a_1,~~~\a_6=\a_6,~~~\a_2=\a_3,~~~\a_4=\a_5
~~\label{rel-5-2}\eea
In other words, using the relabeling property of   ${\cal P}(1,5)$ we have reduced six unknown
variables to four unknown variables.

Now we discuss the implication of relabeling property coming from permutation ${\cal P}(4,5)$. On
the  one hand we have
\bea \tau_{12345}\to \tau_{12354}& = & \a_1 n_{12354}+ \a_2
n_{12534}+\a_3 n_{13254}+\a_4 n_{13524}+ \a_5 n_{15234}+\a_6
n_{15324}\nn
& = & n_{12345}(-\a_1-\a_2-\a_5)+
n_{12435}(\a_2+\a_5+\a_6)+n_{13245}(-\a_3-\a_4-\a_6)\nn & & +
n_{13425}(\a_4+\a_5+\a_6)+n_{14235}(-\a_6)+ n_{14325} (-\a_5)\eea
%
The first line of the equation derives from relabeling,
which subsequently produces the second line using \eref{DDM-relations-1-1}.
On the other hand we have
\bea \tau_{12345}\to \tau_{12354}& = &
-\tau_{12345}-\tau_{12435}-\tau_{14235}\nn
& = & n_{12345}(-\a_1-\a_2-\a_4)+
n_{12435}(-\a_2-\a_1-\a_3)+n_{13245}(-\a_3-\a_5-\a_6)\nn & +&
n_{13425}(-\a_4-\a_6-\a_5)+n_{14235}(-\a_5-\a_3-\a_1)+ n_{14325}
(-\a_4-\a_6-\a_2)\eea
where in the first line we have used   KK-relation
 and in the
second line we have used the expansion of $\tau$'s into $n_\a$. Comparing
the above
two results, we obtain the following equations
\bea & &
(-\a_1-\a_2-\a_5)=(-\a_1-\a_2-\a_4),~~~\a_2+\a_5+\a_6=(-\a_2-\a_1-\a_3),\nn
& &  (-\a_3-\a_4-\a_6)=(-\a_3-\a_5-\a_6),~~~~\a_4+\a_5+\a_6
=(-\a_4-\a_6-\a_5)\nn & & -\a_6=(-\a_5-\a_3-\a_1),~~~
-\a_5=(-\a_4-\a_6-\a_2)~~~\label{rel-5-3}\eea
Combining \eref{rel-5-2} and \eref{rel-5-3} we   find
\bea \a_2=\a_3=-{2\over 5}\a_1,~~~\a_6={2\over
5}\a_1,~~~\a_4=\a_5=-{1\over 5}\a_1\eea
If we put $\a_1={1\over 4}$ back, we do reproduce the result
\eref{tau5p-12345}.

The above example can be generalized to arbitrary number of legs. The point
of this second approach is that if we fully use the potential of
relabeling property, i.e, the relabeling properties of
all generators of permutation group $S_n$, we will be able to determine
expressions  of all $\tau$'s over $(n-2)!$ BCJ
numerator  $n_{1\sigma n}$ up to an overall factor (for
example, $\a_1$ in above example). It is crucial to notice that in the second approach,
we have used only cyclic symmetry \eref{cyclic-symmetry} and KK-relations \eqref{KK-relation}
among $\tau$'s, but not the relation \eqref{n-tau-relation}, which relates $\tau$ with
BCJ numerators $n_\a$. In other words, relabeling property, cyclic symmetry plus KK-relations
for $\tau$'s {\bf have uniquely determined} the expression of $\tau$
in terms of BCJ numerators $n_{\a}$ up
to an overall constant!

To determine the overall factor, relation such as \eref{n-tau-relation} enters the game.
However, based on our examples in section 4, we found the following
pattern of the expansion of $\tau_{123...(n-1)n}$:
%
\bea \tau_{123} &= & {1\over 2} n_{123} \nn
\tau_{1234} & = & {1\over 3} n_{1234}+... \nn
\tau_{12345} & = & {1\over 4} n_{12345}+...\nn
\tau_{123456} & = & {1\over 5} n_{123456}+...\nn
\tau_{1234567} & = & {1\over 6} n_{1234567}+...~~~\label{obser}\eea
We believe that this pattern is right although we can not prove it
at the moment. If we
accept this as an assumption, we find that \eref{n-tau-relation}
is not needed
anymore.

Now we can see the difference between the second approach and the algorithm
presented in section 3. For algorithm in section 3, the equation \eref{n-tau-relation}
is crucial. However, in the second approach, the relabeling property is
crucial. In fact,  using the relabeling property, plus cyclic symmetry and
KK-relation, the equation \eref{n-tau-relation} can be derived
if we use our observation \eref{obser}.

\subsection{The implication of permutation ${\cal P}(1,n)$}

From our previous discussions for the second approach, we see that all nontrivial equations
for $(n-2)!$ expansion coefficients of $\tau_{123...(n-1)n}$ are
given by relabeling properties coming from permutations ${\cal P}(1,n)$ and
${\cal P}(n-1,n)$. These equations coming
from permutation  ${\cal P}(n-1, n)$ will be complicated to write down. However, these equations
coming from permutation
${\cal P}(1,n)$ are very simple and we will present them in this subsection.

Let us start with the expansion
\bea \tau_{123..(n-1)n}= \sum_{\sigma\in S_{n-2}} c_\sigma
n_{1\sigma n} \eea
with $(n-2)!$ coefficients $c_\sigma$. The relabeling by permutation  $
{\cal P}(1,n)$, i.e,
 $1\leftrightarrow n$, leads to the expression
\bea \tau_{n23...(n-1) 1} =\sum_\sigma c_\sigma n_{n\sigma 1}~~~~\label{tau-ex-1}\eea
Using the cyclic symmetry and KK-relation for $\tau_{n23...(n-1) 1}$ we arrive
another expression of $\tau_{n23...(n-1) 1}$
\bea \tau_{n23...(n-1) 1}= \tau_{1n23...(n-1) }=(-)^n
\tau_{1(n-1)(n-2)...32n}=(-)^n \sum_\sigma c_\sigma n_{1\W{ \cal P}(\sigma)
n}~~~~\label{tau-ex-2}\eea
where at the last equation we have used the fact that the expansion
of $\tau_{1(n-1)(n-2)...32n}$ can be obtained from the expansion of
$\tau_{123..(n-1)n}$ by following permutation
\bea \W {\cal P}\equiv \left\{ \begin{array}{ll}(2,n-1)(3,n-2)...(n/2,n/2+1) & n=even\\
(2,n-1)(3,n-2)...((n-1)/2,(n+1)/2+1)~~~~ &n=odd \end{array}\right.\eea
Identifying two different expressions \eref{tau-ex-1} and \eref{tau-ex-2} we have
\bea (-)^n \sum_\sigma c_\sigma n_{1\W {\cal P}(\sigma) n}=\sum_{\W\sigma}
c_{\W\sigma} n_{n\W\sigma 1} =(-)^n \sum_{\W\sigma} c_{\W\sigma}
n_{1(\W\sigma)^T n}~~~\label{tau-coeff-1}\eea
where we have used $(\W\sigma)^T$ to denote reversing the ordering of the list $\W\sigma$.
Since each BCJ numerator $n_\a$ is independent, to have identity \eref{tau-coeff-1},
coefficient of each $n_\a$ must be the same at both sides. Identifying
$n_{1\W {\cal P}(\sigma) n}=n_{1(\W\sigma)^T n}$ for given pair of $\sigma,\W\sigma$, we find
following result: when two orderings $\sigma$ and $\W\sigma$ are related to each other
by $\W{\cal P}(\sigma)= (\W\sigma)^T$, their coefficients must be the same, i.e., we will have
\bea  c_{\W\sigma}=
c_\sigma~~\label{P1n-result}\eea
or more explicitly
\bea c_{1\sigma n}= c_{1 (\W {\cal P}(\sigma))^T
n},~~~~\forall \sigma\in S_{n-2}~~~~\label{P1n-result-1}\eea
Results \eref{P1n-result-1} are  equations coming from relabeling property
of permutation
${\cal P}(1n)$. It is easy to check it with the explicit results given in
section 4.

From \eref{P1n-result-1} we see that there are two special orderings
which are singlet under about transformation. They are
$c_{1234...(n-1)n}$ and $c_{1(n-1)(n-2)...32n}$. In fact, all
coefficients will organize themselves to orbits with one or two
elements under the mapping \eref{P1n-result-1}.

\section{Conclusion}
\label{sec:concln}
In this paper, we have constructed the dual-trace decomposition for Yang-Mills tree amplitudes from
 kinematic numerators.
By imposing cyclic symmetry and KK relation, and the relation between $\tau$'s and BCJ numerators in KK-basis,
we find solutions of $\tau$'s as   linear combinations of BCJ numerators.
The dual-trace factors solved in this way are related to each other by relabeling. Thus we can get any dual trace factor
by relabeling a single $\tau$ with given permutation.
We find that we can also turn things around and start with the relabeling property
to fix the dual-trace factors $\tau$.

\appendix

\section{The relabeling of permutation $P_{1i}$}

To complete the proof of natural relabeling property, we need to prove that
\eref{G-con-rel} is satisfied for the permutation ${\cal P}_{1i}=(1,i)$. In this Appendix
we
provide a detailed proof. We note that equation \eref{G-con-rel} can be
rewritten as
\bea & &
\sum_{\{\W\rho\},\{\W\sigma\}}P(i,\{\rho'\},1,\{\sigma'\},n|1,\{\W\rho\},i,\{\W\sigma\},n)G
(1,\{\W\rho\},i,\{\W\sigma\},n|1,\{\rho\},i,\{\sigma\},n)\nn & =
 & \sum_{\{\rho''\},\{\sigma''\}}G(i,\{\rho'\},1,\{\sigma'\},n|
i,\{\rho''\},1,\{\sigma''\},n)Q(i,\{\rho''\},1,\{\sigma''\},n|1,\{\rho\},i,\{\sigma\},n)
~~~\label{re-prop2}\eea
%
We introduce curly brackets to  emphasize that, for example,
$\{\rho\}$ stands for a list of elements $\rho_{1}$, $\rho_{2}$, $\dots$
(which can be empty as well), and that the ordering of the list
matters. We break the proof into three steps discussed below:

~\\{\bf Step-1: Finding  explicit expressions of  elements of $P$
and $Q$ matrices}

Under the
permutation ${\cal P}(1,i)$,
 a KK-basis numerator characterized by the fixed pair $(1,n)$
 changes to a
KK-basis numerator characterized by  $(i,n)$,
and we need the collaboration of
Jacobi identity, antisymmetry to derive the matrix of $P$, and  KK
relation together with cyclic symmetry to derive the matrix of $Q$
in $(1,n)$ basis.

Let us consider the matrix $P$ first. Using the Jacobi identity and
antisymmetry, we have
\bea n_{i,\{\rho\},1,\{\sigma\},n}=\Sl_{\{\rho\}\rightarrow
\{\alpha\}\{\beta\}}(-1)^{n_{\alpha}+1}n_{1,\{\alpha\}^T,i,\{\beta\},\{\sigma\},n}.
\eea
where $\Sl_{\{\rho\}\rightarrow \{\alpha\}\{\beta\}}$ means summing
over all possible splittings of set $\{\rho\}$ to two subsets
$\{\a\},\{\b\}$ with their relative orderings kept. As remarked at
the beginning of section \ref{sec:const-ntr}, this is equivalent to
summing over all the ordered sets $\{\a\},\{\b\}$ satisfying the
condition $\{\rho\}\in OP(\{\alpha\} \bigcup\{\beta\})$, yet the
advantage of regarding this process as a splitting instead of as
imposing a constraint will become obvious shortly as the complexity
increases. It is straightforward then, to read off the elements of
matrix $P$ from the above equation
\bea
P(i,\{\rho\},1,\{\sigma\},n|1,\{\alpha\}^T,i,\{\beta\},
         \{\sigma\},n)=
    \Biggl\{
      \begin{array}{cc}
         (-1)^{n_{\alpha}+1} & (\text{if $\{\rho\}$
         can splits into $\{\alpha\},\{\beta\}$}) \\
        0 &\text{ Otherwise} \\
      \end{array}.
    ~~\label{P-element}
\eea

Next we consider the matrix $Q$. For
$\tau_{i,\{\rho\},1,\{\sigma\},n}$ we can use  cyclic symmetry and
KK relation
\bea
\tau_{i,\{\rho\},1,\{\sigma\},n}=\tau_{1,\{\sigma\},n,i,\{\rho\}}=\Sl_{\{\delta\}\in
OP(\{\sigma\}\bigcup\{\rho^T,i\})}(-1)^{n_{\rho}+1}\tau_{1,\{\delta\},n}.
\eea
thus elements of $Q$ can be read off
\bea
 Q(i,\{\rho\},1,\{\sigma\},n|1,\{\delta\},n)=
 \Biggl\{
  \begin{array}{cc}
   (-1)^{n_{\rho}+1} &
    \text{if $\{\delta\}\in OP(\{\sigma\}\bigcup\{\rho^T,i\})$} \\
    0 & \text{Otherwise} \\
  \end{array}.~~\label{Q-element}
\eea
Because the ordering  $\{\rho^T,i\}$ in \eref{Q-element},
nonzero elements of $Q$ must  have the form
\bea
Q(i,\{\alpha^T\},1,\{\beta\},\{\sigma\},n|1,\{\rho\},i,\{\sigma\},n)=(-1)^{n_{\alpha}+1},~~~
\text{if $\{\rho\}\in   OP(\{\alpha\}\bigcup
\{\beta\})$.}~~\label{Q-element-1}\eea

Plugging the newly obtained
explicit expression of matrix elements of $P$,  the matrix $PG$ in
 \eqref{re-prop2} can be expressed as
\bea &&\Sl_{\{\rho'\}\rightarrow
\{\W\alpha\},\{\W\beta\}}(-1)^{n_{\W\alpha}+1}G_{1,n}(1,\{\W\alpha^T\},i,\{\W\beta\},\{\sigma'\},
n|1,\{\rho\},i,\{\sigma\},n).~~~\label{A7}\eea
Similarly, the matrix  $GQ$ in \eqref{re-prop2} is given by
\bea \Sl_{\{\rho\}\rightarrow \{\alpha\},\{\beta\}}
G_{i,n}(i,\{\rho'\},1,\{\sigma'\},n|i,\{\alpha^T\},1,\{\beta\},\{\sigma\},n)(-1)^{n_{\alpha}+1}.
~~~\label{A8}\eea
To compare \eref{A7} and \eref{A8}, using the relabeling invariant
property of matrix $G$, we can exchange the positions of $1$ and $i$ in
\eref{A8} and this  expression becomes
\bea \Sl_{\{\rho\}\rightarrow \{\alpha\},\{\beta\}}
G_{1,n}(1,\{\rho'\},i,\{\sigma'\},n|1,\{\alpha^T\},i,\{\beta\},\{\sigma\},n)(-1)^{n_{\alpha}+1}.
\eea
Finally the consistency condition \eqref{re-prop2} can be rewritten
as
\bea &&\Sl_{\{\rho'\}\rightarrow
\{\alpha'\},\{\beta'\}}(-1)^{n_{\alpha'}+1}G_{1,n}(1,\{\alpha'^T\},i,
\{\beta'\},\{\sigma'\},n|1,\{\rho\},i,\{\sigma\},n)\nn
&=&\Sl_{\{\rho\}\rightarrow \{\alpha\},\{\beta\}}
~~~(-1)^{n_{\alpha}+1}G_{1,n}(1,\{\rho'\},i,\{\sigma'\},n|1,\{\alpha^T\},i,\{\beta\},\{\sigma\},n)
.\label{re-prop2-1} \eea
%

We need to show the
sums at  both sides of \eref{re-prop2-1} give the same result.

~\\{\bf Step-2: Further simplification by  properties of $G$}

Before finally deriving a proof let us further simplify the condition
\eqref{re-prop2}. We notice that $G_{1,n}(\alpha|\beta)$ can be
written by the property \eqref{G-delta-1} as
\bea
G_{1,n}(\sigma'|\sigma)=\Sl_{s'\in\{{\cal S}(\sigma')\}}\Sl_{s\in\{{\cal S}(\sigma)\}}\delta(s'|s),~~\label{G-delta} \eea
where to simplify the notation, we have used ${\cal S}(\sigma')$ to denote the set of all possible splittings
of the set $\{\sigma'\}$ into two subsets, for example $\{\a'\},\{\b'\}$, with relative ordering
kept, and the sum is taken over all elements of the set ${\cal S}(\sigma')$. The delta-function is defined as
\bea \delta(s'|s)=\Biggl\{
  \begin{array}{cc}
    1 & (s'=s) \\
    0 & \text{Otherwise} \\
  \end{array}.
\eea
where $s'=s$ means that both
$\{\a'\}=\{\a\}$ and $\{\b'\}=\{\b\}$ if $\{\sigma'\}$ is split to $\{\a'\},\{\b'\}$ and $\{\sigma\}$
is split to $\{\a\},\{\b\}$.

Substituting \eqref{G-delta} into \eqref{re-prop2-1}, we get
\bea \Sl_{\{\rho'\}\rightarrow
\{\alpha'\},\{\beta'\}}(-1)^{n_{\alpha'}}\left[\Sl_{s',s}\delta(s'|s)\right]=\Sl_{\{\rho\}\rightarrow
\{\alpha\},\{\beta\}}(-1)^{n_{\alpha}}\left[\Sl_{s',s}\delta(s'|s)\right],\nn
\eea where we have summed over $s'\in\{{\cal
S}(\{\alpha'^T\},i,\{\beta'\},\{\sigma'\})\}$ and $s\in\{{\cal
S}(\{\rho\},i,\{\sigma\})\}$ on the L.H.S, while we have summed over
$s'\in\{{\cal S}(\{\rho'\},i,\{\sigma'\})\}$ and $s\in\{{\cal
S}(\{\alpha^T\},i,\{\beta\},\{\sigma\})\}$ on the R.H.S. The above
equation can be further rearranged into
\bea &&\Sl_{s\in\{{\cal
S}(\{\rho\},i,\{\sigma\})\}}\left[\Sl_{\{\rho'\}\rightarrow
\{\alpha'\},\{\beta'\}}(-1)^{n_{\alpha'}}\Sl_{s'\in\{{\cal
S}(\{\alpha'^T\},i,\{\beta'\},\{\sigma'\})\}}\delta(s'|s)\right]\nn
&=&\Sl_{s'\in\{{\cal
S}(\{\rho'\},i,\{\sigma'\})\}}\left[\Sl_{\{\rho\}\rightarrow
\{\alpha\},\{\beta\}}(-1)^{n_{\alpha}}\Sl_{s\in\{{\cal
S}(\{\alpha^T\},i,\{\beta\},\{\sigma\})\}}\delta(s'|s)\right].~~\label{re-prop2-2}\nn
\eea
For a given splitting $s$ the sum in the brackets on the L.H.S. has
a useful property
\bea &&\Sl_{\{\rho'\}\rightarrow
\{\alpha'\},\{\beta'\}}(-1)^{n_{\alpha'}}\Sl_{s'\in\{{\cal
S}(\{\alpha'^T\},i,\{\beta'\},\{\sigma'\})\}}\delta(s'|s)\nn
&=&\Sl_{\{\rho'\}\rightarrow
\{\alpha'\},\{\beta'\}}(-1)^{n_{\alpha'}}\Sl_{s'\in\{{\cal
S}(\{\sigma'\}\rightarrow~\{\sigma'_L\},\{\sigma'_R\})\}}
\delta(\{\{\alpha'^T\},i,\{\beta'\},\{\sigma'_L\}\},\{\sigma'_R\}|s).~~\label{G-delta-prop}\nn
\eea
For a given splitting $s'$ the sum in the brackets on the R.H.S. has
a similar property. The meaning of \eref{G-delta-prop} is that for
all possible splittings only those with $\{\alpha'^T\},i,\{\beta'\}$
belonging to the same subset contribute\footnote{One may notice the splitting can be either $\{\sigma_1,i,\sigma_2\},\{\sigma_3\}$ or $\{\sigma_3\},\{\sigma_1,i,\sigma_2\}$.
Since there is no difference for  following discussions between the two kinds of splittings, we just need to deal with only the first kind.  }.

Before giving a general proof of the above property
\eref{G-delta-prop}, let us have a look at some examples.
\begin{itemize}
\item (1) For the case $n_{\rho'}=1$, i.e., there is only one
element in the set $\rho'$, there are two possible splittings:
$\{\a',\b'\}=\{ \{~\}, \{\rho_1'\}\}/\{  \{\rho_1'\}, \{~\}\}$. For
the case $\a'=\{\rho_1'\}$, the splitting of $\{\rho_1',i,
\sigma'\}$ contains two possibilities: either $\rho_1'$ and $i$ belong to the same
subset or   to different subsets. For the case
$\b'=\{\rho_1'\}$, the splitting of $\{i,\rho_1', \sigma'\}$ also
contains two possibilities: either $\rho_1',i$ belong to the same subset or
 to different subsets. Putting all these together,
 the L.H.S. of \eqref{G-delta-prop} reads
\bea
&&\Sl_{\{\sigma'\}\rightarrow\{\sigma_L\},\{\sigma_R\}}\Biggl[-\delta
\left(\{\rho'_1,i,\sigma'_L\},\{\sigma'_R\}|s\right) -\delta\left(
\{\rho'_1,\sigma'_L\},\{i,\sigma'_R\}|s\right) +\delta\left(
\{i,\rho'_1,\sigma'_L\},\{\sigma'_R\}|s\right)\nn
&&+\delta\left(\{i,\sigma'_L\},\{\rho'_1,\sigma'_R\}|s\right)\Biggr]
\eea

After summing over all  splittings $\{\sigma'\}\rightarrow
\{\sigma'\}_L,\{\sigma'\}_R$, the second term and the fourth term
 cancel each other and only two terms are left
\bea \Sl_{\{\sigma'\}\rightarrow\{\sigma_L\},\{\sigma_R\}}\Biggl[-\delta
\left(\{\rho'_1,i,\sigma'_L\},\{\sigma'_R\}|s\right) +\delta\left(
\{i,\rho'_1,\sigma'_L\},\{\sigma'_R\}|s\right)\Biggr]. \eea
which is  just  the R.H.S. of \eqref{G-delta-prop} when
$n_{\rho'}=1$.

\item (2) For the case $n_{\rho'}=2$, similar consideration as the case
$n_{\rho'}=1$ gives the L.H.S.:
\bea
&&\Sl_{\{\sigma'\}\rightarrow\{\sigma'_L\},\{\sigma'_R\}}\Biggl\{\Biggl[
\delta\left(\{\rho'_2,\rho'_1,i,\sigma'_L\},\{\sigma'_R\}|s\right)
+\delta\left(
\{\rho'_2,i,\sigma'_L\},\{\rho'_1,\sigma'_R\}|s\right)\nn &&
+\delta\left(
\{\rho'_1,i,\sigma'_L\},\{\rho'_2,\sigma'_R\}|s\right)
+\delta\left(\{\rho'_2,\rho'_1,\sigma'_L\},\{i,\sigma'_R\}|s\right)\Biggr]\nn
&&-\Biggl[\delta\left(\{\rho'_2,i,\rho'_1,\sigma'_L\},\{\sigma'_R\}|s\right)
+\delta\left(
\{\rho'_2,i,\sigma'_L\},\{\rho'_1,\sigma'_R\}|s\right)\nn
&&+\delta\left(
\{\rho'_2,\rho'_1,\sigma'_L\},\{i,\sigma'_R\}|s\right)
+\delta\left(\{\rho'_1,i,\sigma'_L\},\{\rho'_2,\sigma'_R\}|s\right)\Biggr]\nn
&&-\Biggl[\delta\left(\{\rho'_1,i,\rho'_2,\sigma'_L\},\{\sigma'_R\}|s\right)
+\delta\left(
\{\rho'_1,i,\sigma'_L\},\{\rho'_2,\sigma'_R\}|s\right)\nn
&&+\delta\left(
\{\rho'_1,\rho'_2,\sigma'_L\},\{i,\sigma'_R\}|s\right)
+\delta\left(\{\rho'_2,i,\sigma'_L\},\{\rho'_1,\sigma'_R\}|s\right)\Biggr]\nn
&&+\Biggl[\delta\left(\{i,\rho'_1,\rho'_2,\sigma'_L\},\{\sigma'_R\}|s\right)
+\delta\left(
\{i,\rho'_1,\sigma'_L\},\{\rho'_2,\sigma'_R\}|s\right)\nn &&
+\delta\left(
\{i,\rho'_2,\sigma'_L\},\{\rho'_1,\sigma'_R\}|s\right)
+\delta\left(\{\rho'_1,\rho'_2,\sigma'_L\},\{i,\sigma'_R\}|s\right)\Biggr]\Biggr\}
\eea
Again, after summing over all splittings of $\{\sigma'\}$,  terms
cancel each other and we are left with
\bea &&\Sl_{\{\sigma'\}\rightarrow\{\sigma'_L\},\{\sigma'_R\}}\Biggl[
\delta\left(\{\rho'_2,\rho'_1,i,\sigma'_L\},\{\sigma'_R\}|s\right)
-\delta\left(\{\rho'_2,i,\rho'_1,\sigma'_L\},\{\sigma'_R\}|s\right)\nn
&&-\delta\left(\{\rho'_1,i,\rho'_2,\sigma'_L\},\{\sigma'_R\}|s\right)
+\delta\left(\{i,\rho'_1,\rho'_2,\sigma'_L\},\{\sigma'_R\}|s\right)
\Biggr], \eea
which is just the R.H.S. of \eqref{G-delta-prop} in the case of
$n_{\rho'}=2$.

\item (3) Similar calculation has  been done   for the case of
$n_{\rho'}=3$,
which we neglect here, since the manipulations are quite similar to
the examples shown.
\end{itemize}

Above examples give the idea of proof. For the case with
$n_{\rho'}=r$ after the splitting $\{\rho'\}\rightarrow
\{\alpha'\},\{\beta'\}$ with $n_{\a'}=s$ and $n_{\b'}=r-s$, we
need to sum over all splittings of $\{\a',i,\b',\sigma'\}$. In
general, the splitting will be $\{\a'_1,i,\b'_1,\sigma'_1\},
\{\a'_2,\b'_2,\sigma'_2\}$. Among these two subsets, unlike the
subset
 $\{\a'_1,i,\b'_1,\sigma'_1\}$ where $i$ seperates the $\a'$ part from
 $\b'$ part, the subset
$\{\a'_2,\b'_2,\sigma'_2\}$ can come from different splittings of
$\{\rho'\}\rightarrow \{\alpha'\},\{\beta'\}$. More
 explicitly, the last element of $\a'_2$ can be considered as the
 first element of $\b_2''$. Thus when we put the factor $(-)^{n_{\a'}}$
 back, the term coming from $n_{\a'}=s$ will cancel with the term
 coming from $n_{\a''}=s+1$. Because
this kind of cancelations, only splittings  with all  elements of
$\{\alpha^T\},i,\{\beta\}$ in the same ordered subset contribute.

Having established \eqref{G-delta-prop}, \eref{re-prop2-2} can be
rewritten as
{\bea \Sl_{s\in\{{\cal
S}(\{\rho\},i,\{\sigma\})\}}\Biggl[\Sl_{\{\rho'\}\rightarrow
\{\alpha'\},\{\beta'\}}(-1)^{n_{\alpha'}}\Sl_{{\cal
S}(\{\sigma'\}\rightarrow~\{\sigma'_L\},\{\sigma'_R\})}
\delta(\{\{\alpha'^T\},i,\{\beta'\},\{\sigma'_L\}\},\{\sigma'_R\}|s)\Biggr]~~\label{re-prop2-3}\\
=\Sl_{s'\in\{{\cal
S}(\{\{\rho'\},i,\{\sigma'\}\})\}}\Biggl[\Sl_{\{\rho\}\rightarrow
\{\alpha\},\{\beta\}}(-1)^{n_{\alpha}}\Sl_{{\cal
S}(\{\sigma\}\rightarrow~\{\sigma_L\},\{\sigma_R\})}
\delta(\{\{\alpha^T\},i,\{\beta\},\{\sigma_L\}\},\{\sigma_R\}|s')\Biggr].\nonumber
\eea }

~\\{\bf Step-3: Proving the relabeling properties $PG=GQ$ via
\eqref{re-prop2-3}}

From step-1 and step-2, we have rewritten the relabeling property
$PG=GQ$ to the form   \eqref{re-prop2-3}. Now let us prove
\eqref{re-prop2-3} by considering various configurations:

\begin{itemize}

\item (1) {\bf Both $\{\rho\}$ and $\{\rho'\}$ are
empty: } In this case,  \eref{re-prop2-3} becomes
\bea &&\Sl_{{\cal S}(\{\sigma\}\rightarrow
\{\sigma_L\},\{\sigma_R\})}\Biggl[\Sl_{{\cal
S}(\{\sigma'\}\rightarrow~\{\sigma'_L\},\{\sigma'_R\})}
\delta(\{\sigma'_L\},\{\sigma'_R\}|\{\sigma_L\},\{\sigma_R\})\Biggr]\nn
&=&\Sl_{{\cal S}(\{\sigma'\}\rightarrow
\{\sigma'_L\},\{\sigma'_R\})}\Biggl[\Sl_{{\cal
S}(\{\sigma\}\rightarrow~\{\sigma_L\},\{\sigma_R\})}
\delta(\{\sigma_L\},\{\sigma_R\}|\{\sigma'_L\},\{\sigma'_R\})\Biggr],\nn
\eea
which is trivially true.

\item (2) {\bf Only one of $\{\rho\}$ and $\{\rho'\}$ is empty:}
Assuming $\{\rho\}$ is empty,  possible nontrivial terms on the L.H.S.
of \eqref{re-prop2-3} is given as
\bea &&\Sl_{\{i,\sigma\}\rightarrow
\{i,\sigma_L\},\{\sigma_R\}}\Sl_{\{\sigma'\}
\rightarrow~\{\sigma'_L\},\{\sigma'_R\}}\delta(\{i,\rho',\sigma'_L\},\{\sigma'_R\}|
\{i,\sigma_L\},\{\sigma_R\})\nn
&=&\Sl_{\{\sigma\}\rightarrow
\{\sigma_L\},\{\sigma_R\}}\Sl_{\{\sigma'\}\rightarrow~\{\sigma'_L\},\{\sigma'_R\}}\delta(\{\rho',\sigma'_L\}|\{\sigma_L\})\delta(\{\sigma'_R\}|
,\{\sigma_R\}). \eea
where we have used the fact that since the splitting of $s$ has element $i$
at the first position, nonzero contribution requires the splitting
of $\{\rho'\}$ to be $\{\a'\}$ empty.

For the  R.H.S. contribution is
\bea &&\Sl_{s'\in {\cal
S}(\{\rho',i,\sigma'\})}\Sl_{\{\sigma\}\rightarrow~\{\sigma_L\},\{\sigma_R\}}
\delta(\{i,\sigma_L\},\{\sigma_R\}|s')\nn
&=&\Sl_{\{\sigma'\}\rightarrow
\{\sigma'_L\},\{\sigma'_R\}}\Sl_{\{\sigma\}\rightarrow~\{\sigma_L\},\{\sigma_R\}}
\delta(\{i,\sigma_L\},\{\sigma_R\}|\{i,\sigma'_L\},\{\rho',\sigma'_R\})\nn
&=&\Sl_{\{\sigma\}\rightarrow
\{\sigma_L\},\{\sigma_R\}}\Sl_{\{\sigma'\}\rightarrow~\{\sigma'_L\},\{\sigma'_R\}}\delta(\{\rho',\sigma'_L\}|\{\sigma_L\})\delta(\{\sigma'_R\}|
,\{\sigma_R\}). \eea
where in the second line, to have nonzero result, the splitting of
$s'$ is that $\{\rho'\}$ and $i$ belong to different subsets. Thus in
this case, the relabeling property is also satisfied.

\item (3) {\bf Both $\{\rho'\}$ and $\{\rho\}$ are nonempty and they
have no element in common:} Let us consider the L.H.S. of
\eqref{re-prop2-3} first.  Since  $\{\rho\}$ and $\{\rho'\}$
have no element in common,
to have nonzero contribution, we  must have $\{\a'\}$ empty in
the splitting of $\{\rho'\}$ and $\{\rho, i\}$ belong to different subsets
in the splitting of $s$. Thus the L.H.S. of \eqref{re-prop2-3} is
given as
\bea &&\Sl_{\{\sigma\}\rightarrow
\{\sigma_L\},\{\sigma_R\}}\Sl_{\{\sigma'\}\rightarrow
\{\sigma'_L\},\{\sigma'_R\}}\delta(\{i,\rho',\sigma'_L\},\{\sigma'_R\}|\{i,\sigma_L\},\{\rho,\sigma_R\})\nn
&=&\Sl_{\{\sigma\}\rightarrow
\{\sigma_L\},\{\sigma_R\}}\Sl_{\{\sigma'\}\rightarrow
\{\sigma'_L\},\{\sigma'_R\}}
\delta(\{i,\rho',\sigma'_L\}|\{i,\sigma_L\})\delta(\{\sigma'_R\}|\{\rho,\sigma_R\}).
\eea
For the  R.H.S., we need the $\{\a\}$ to be empty in the splitting of
$\{\rho\}$ and $\{\rho', i\}$ belong to different subsets in the splitting
of $s'$, thus  we get
\bea \Sl_{\{\sigma'\}\rightarrow
\{\sigma'_L\},\{\sigma'_R\}}\Sl_{\{\sigma\}\rightarrow
\{\sigma_L\},\{\sigma_R\}}
\delta(\{i,\rho,\sigma_L\}|\{i,\sigma'_L\})\delta(\{\sigma_R\}|\{\rho,\sigma'_R\})
\eea
Thus the relabeling property in this case is satisfied.

\item (4) {\bf Both $\{\rho\}$ and $\{\rho'\}$ are nonempty and they share
common elements:} This is a most general case. The L.H.S. of
\eqref{re-prop2-3} is
\bea &&\Sl_{\{\rho\}\rightarrow
\{\alpha\},\{\beta\}}\Sl_{\{\sigma\}\rightarrow
\{\sigma_L\},\{\sigma_R\}}\Biggl[\Sl_{\{\rho'\}\rightarrow
\{\alpha'\},\{\beta'\}}(-1)^{n_{\alpha'}}\nn
&&\times\Sl_{\{\sigma'\}\rightarrow~\{\sigma'_L\},\{\sigma'_R\}}
\delta(\{\alpha'^T,i,\beta',\sigma'_L\},\{\sigma'_R\}|\{\alpha,i,\sigma_L\},\{\beta,\sigma_R\})\Biggr]\nn
&=&\Sl_{\{\rho\}\rightarrow
\{\alpha\},\{\beta\}}\Sl_{\{\sigma\}\rightarrow
\{\sigma_L\},\{\sigma_R\}}\Sl_{\{\rho'\}\rightarrow
\{\alpha'\},\{\beta'\}}\Sl_{\{\sigma'\}\rightarrow~\{\sigma'_L\},\{\sigma'_R\}}\Biggl[(-1)^{n_{\alpha'}}\nn
&&\times\delta(\{\alpha'^T\}|\{\alpha\})\delta(\{\beta',\sigma'_L\}|\{\sigma_L\})\delta(\{\sigma'_R\}|\{\beta,\sigma_R\})\Biggr],
\eea Similarly, the R.H.S. of \eqref{re-prop2-3} can be written as
\bea &&\Sl_{\{\rho'\}\rightarrow
\{\alpha'\},\{\beta'\}}\Sl_{\{\sigma'\}\rightarrow
\{\sigma'_L\},\{\sigma'_R\}}\Sl_{\{\rho\}\rightarrow
\{\alpha\},\{\beta\}}\Sl_{\{\sigma\}\rightarrow~\{\sigma_L\},\{\sigma_R\}}\Biggl[(-1)^{n_{\alpha}}\nn
&&\times\delta(\{\alpha^T\}|\{\alpha'\})\delta(\{\beta,\sigma_L\}|\{\sigma'_L\})\delta(\{\sigma_R\}|\{\beta',\sigma'_R\})\Biggr].
\eea
Changing $\a\to \a'$ etc,  and the L.H.S.  equals   the R.H.S. in this
case.

\end{itemize}

Taking all the above
possibilities into account, the property \eqref{re-prop2-3} is satisfied for all
cases, and we see that the relabeling condition \eqref{re-prop2} is satisfied
for permutation ${\cal P}(1,i)$.

\subsection*{Acknowledgements}
Y. J. Du would like to thank Prof. Yong-Shi Wu for helpful suggestions. Y. J. Du is
supported in part by the NSF of China Grant No.11105118.  CF would
 like to acknowledge
the support from National Science Council, 50 billions project of
Ministry of Education and National Center for Theoretical Science,
Taiwan, Republic of China as well as the support from S.T. Yau
center of National Chiao Tung University. B.F is supported, in part,
by fund from Qiu-Shi and Chinese NSF funding under contract
No.11031005, No.11135006, No. 11125523.


\end{document}